\documentclass[12pt,fleqn,a4paper]{article} 
\usepackage{epsfig,graphics,graphicx}
\usepackage{clshan-math,clshan-dm-dd}
\def\lsim{\:\raisebox{-0.5ex}{$\stackrel{\textstyle<}{\sim}$}\:}
\def\gsim{\:\raisebox{-0.5ex}{$\stackrel{\textstyle>}{\sim}$}\:}
\begin{document}
\renewcommand{\thefootnote}{\fnsymbol{footnote}}

\thispagestyle{empty}
\begin{flushright}
KIAS--P08026 \\
March 2008
\end{flushright}
\begin{center}
{\Large\bf
 Model--Independent Determination of the WIMP Mass      \\  \vspace{0.2cm}
  from Direct Dark Matter Detection Data}          \\
\vspace*{0.7cm}
 {\sc Manuel Drees}$^{1,2}$ and {\sc Chung-Lin Shan}$^1$ \\
\vspace*{0.3cm}
${}^1$ {\it Physikalisches Inst. der Univ. Bonn, Nussallee 12, 53115 Bonn,
 Germany} \\
${}^2$ {\it KIAS, School of Physics, 207--43 Cheongnyangni--dong, Seoul
 130--012, Republic of Korea}
\end{center}
\vspace{1cm}

\begin{abstract}
  Weakly Interacting Massive Particles (WIMPs) are one of the leading
  candidates for Dark Matter. We develop a model--independent method for
  determining the mass $m_\chi$ of the WIMP by using data (i.e., measured
  recoil energies) of direct detection experiments. Our method is independent
  of the as yet unknown WIMP density near the Earth, of the form of the WIMP
  velocity distribution, as well as of the WIMP--nucleus cross section.
  However, it requires positive signals from at least two detectors with
  different target nuclei. In a background--free environment, $m_\chi \sim
  50$ GeV could in principle be determined with an error of $\sim 35\%$ with
  only $2 \times 50$ events; in practice upper and lower limits on the recoil
  energy of signal events, imposed to reduce backgrounds, can increase the
  error. The method also loses precision if $m_\chi$ significantly exceeds the
  mass of the heaviest target nucleus used.
\end{abstract}
\clearpage
\section{Introduction}

First indications for the existence of Dark Matter were already found in the
1930s \cite{evida}.  By now there is strong evidence \cite{evida}-\cite{bullet}
to believe that a large fraction (more than 80\%) of all matter in the
Universe is dark (i.e., interacts at most very weakly with electromagnetic
radiation and ordinary matter). The dominant component of this cosmological Dark Matter must be
due to some yet to be discovered, non--baryonic particles.  Weakly Interacting
Massive Particles (WIMPs) $\chi$ are one of the leading candidates for Dark
Matter.  WIMPs are stable particles which arise in several extensions of the
Standard Model of electroweak interactions. Typically they are presumed to
have masses between 10 GeV and a few TeV and interact with ordinary matter
only weakly (for reviews, see \cite{susydm}).

Currently, the most promising method to detect many different WIMP candidates
is the direct detection of the recoil energy deposited in a low--background
laboratory detector by elastic scattering of ambient WIMPs on the target
nuclei \cite{detaa, detab}. The recoil energy spectrum can be calculated
from an integral over the one--dimensional WIMP velocity distribution
$f_1(v)$, where $v$ is the absolute value of the WIMP velocity in the
laboratory frame. If this function is known, the WIMP mass $m_\chi$ can be
obtained from a one--parameter fit to the normalized recoil spectrum
\cite{green}, once a positive WIMP signal has been found. However, this
introduces a systematic uncertainty which is difficult to control. We remind
the reader that $N-$body simulations of the spatial distribution of Cold Dark
Matter (which includes WIMPs) seem to be at odds with observations at least
in the central region of galaxies \cite{crisis}; this may have ramifications
for $f_1$ as well. On the theory side, several modifications of the standard
``shifted Maxwellian'' distribution \cite{susydm} have been suggested, ranging
from co-- or counter--rotating halos \cite{rotate} to scenarios where the
three--dimensional WIMP velocity distribution gets large contributions from
discrete ``streams'' with (nearly) fixed velocity \cite{sikivie,streams}.

The goal of our work is to develop model--independent methods which allow to
determine $m_\chi$ directly from (future) experimental data.  This builds on
our earlier work \cite{DMDD}, where we showed how to determine (moments of)
$f_1(v)$ from the recoil spectrum in direct WIMP detection experiments. In
this earlier analysis $m_\chi$ had been the only input required; one does not
need to know the WIMP--nucleus scattering cross section, nor the local WIMP
density. The fact that this method will give a result for $f_1$ for any
assumed value of $m_\chi$ already tells us that one will need at least two
different experiments, with different target nuclei, to model--independently
determine the WIMP mass from direct detection experiments. To do so, one
simply requires that the values of a given moment of $f_1$ determined by both
experiments agree. This leads to a simple expression for $m_\chi$, which can
easily be solved analytically; note that each moment can be used. An
additional expression for $m_\chi$ can be derived under the assumption that
the ratio of scattering cross sections on protons and neutrons is known. This
is true e.g., for spin--independent scattering of a supersymmetric neutralino,
which is the perhaps best motivated WIMP candidate \cite{susydm}; in this case
the spin--independent cross section for scattering on a proton is almost the
same as that for scattering on a neutron. Not surprisingly, the best result
obtains by combining measurements of several moments with that derived from
the assumption about the ratio of cross sections.

The remainder of this article is organized as follows. In Sec.~2 we review
briefly the methods for estimating the moments of the velocity distribution
function, paying special attention to experimentally imposed limits on the
range of allowed recoil energies. In Sec.~3 we will present the formalism for
determining the WIMP mass. Numerical results, based on Monte Carlo simulations
of future experiments, will be presented in Sec.~4. We conclude in Sec.~5.
Some technical details of our calculation will be given in an Appendix.

\section{Determining the moments of the velocity distribution of WIMPs}

In this section we review briefly the method for estimating the moments of the
one--dimensional velocity distribution function $f_1$ of WIMPs from the
elastic WIMP--nucleus scattering data. We first discuss the formalism, and
then describe how it can be implemented directly using (simulated, or future
real) data from direct WIMP search experiments.

\subsection{Formalism}

Our analysis starts from the basic expression for the differential rate for
elastic WIMP--nucleus scattering \cite{susydm}:
\beq \label{eqn2101}
\dRdQ = \calA \FQ \intvmin \bfrac{f_1(v)}{v} dv\, .
\eeq
Here $R$ is the direct detection event rate, i.e., the number of events per
unit time and unit mass of detector material, $Q$ is the energy deposited in
the detector, $F(Q)$ is the elastic nuclear form factor, and $v$ is the
absolute value of the WIMP velocity in the laboratory frame. The constant
coefficient $\calA$ is defined as
\beq \label{eqn2102}
\calA \equiv \frac{\rho_0 \sigma_0}{2 \mchi m_{\rm r,N}^2}\, ,
\eeq
where $\rho_0$ is the WIMP density near the Earth and $\sigma_0$ is the total
cross section ignoring the form factor suppression.  The reduced mass $m_{\rm
  r,N}$ is defined as
\beq \label{eqn2103}
m_{\rm r,N} \equiv \frac{\mchi \mN}{\mchi+\mN}\, ,
\eeq
where $\mchi$ is the WIMP mass and $\mN$ that of the target nucleus.  Finally,
$\vmin$ is the minimal incoming velocity of incident WIMPs that can deposit
the energy $Q$ in the detector:
\beq \label{eqn2104}
\vmin = \alpha \sqrt{Q}\, ,
\eeq
where we define
\beq \label{eqn2105}
\alpha \equiv \sfrac{\mN}{2 m_{\rm r,N}^2}\, .
\eeq
 
Eq.(\ref{eqn2101}) can be solved for $f_1$ \cite{DMDD}:
\beq \label{eqn2106} 
f_1(v) = \calN \cbrac{-2 Q \cdot \ddRdQoFQdQ}\Qva\, , 
\eeq 
where the normalization constant $\calN$ is given by
\beq \label{eqn2107} 
\calN = \frac{2}{\alpha} \cbrac{\intz \frac{1}{\sqrt{Q}} \bdRdQoFQ dQ}^{-1}
\, .  
\eeq 
Note that, first,
because $f_1(v)$ in Eq.(\ref{eqn2106}) is the {\em normalized} velocity distribution,
the normalization constant $\cal N$ here is {\em independent} of
the constant coefficient $\cal A$ defined in Eq.(\ref{eqn2102}).
Second, the integral here goes over the entire physically allowed range of
recoil energies, starting at $Q = 0$. The upper limit of the integral has been
written as $\infty$. However, it is usually assumed that the WIMP flux on
Earth is negligible at velocities exceeding the escape velocity \mbox{$v_{\rm esc}
\simeq 700$ km/s}. This leads to a kinematic maximum of the recoil energy,
\beq \label{qmaxkin}
Q_{\rm max,kin} = \frac {v^2_{\rm esc}} {\alpha^2}\, ,
\eeq
where $\alpha$ has been given in Eq.(\ref{eqn2105}). Eq.(\ref{eqn2107}) then
implies
\beq \label{fnorm}
\int_0^\infty f_1(v) \~ dv = \int_0^{v_{\rm esc}} f_1(v) \~ dv = 1\, .
\eeq
Using Eq.(\ref{eqn2106}), the moments of $f_1$ can be expressed as \cite{DMDD}
\beq \label{eqn2108} 
\expv{v^n} = \int_0^{v_{\rm esc}} v^n f_1(v) \~ dv
= \calN \afrac{\alpha^{n+1}}{2} (n+1) I_n\, , 
\eeq
which holds for all $n \geq 0$. Here the integral $I_n$ is given by
\beq \label{eqn2110} 
I_n = \int_0^{Q_{\rm max,kin}} Q^{(n-1)/2} \bdRdQoFQ dQ\, .  
\eeq 
In this notation, Eq.(\ref{eqn2107}) can be re--written as ${\cal N} = 2 /
(\alpha I_0)$. 

The results in Eqs.(\ref{eqn2106}) and (\ref{eqn2108}) depend on the WIMP mass
$m_{\chi}$ only through the coefficient $\alpha$ defined in
Eq.(\ref{eqn2105}). Evidently any (assumed) value of $m_\chi$ will lead to a
well--defined, normalized distribution function $f_1$ when used in
Eq.(\ref{eqn2106}). Hence $m_\chi$ can be extracted from a {\em single} recoil
spectrum {\em only if} one makes some assumptions about the velocity
distribution $f_1(v)$. 

A model--independent determination of $m_\chi$ thus requires that at least two
different recoil spectra, with two different target nuclei, have been
measured. As we will show in detail in the next section, $m_\chi$ can then be
obtained from the requirement that these two spectra lead to the same moments
of $f_1$.

Before coming to that, we have to incorporate the effects of a finite energy
acceptance of the detector. Any real detector will have a certain threshold
energy $Q_{\rm thre}$ below which it cannot register events. Off--line one may
need to impose a cut $Q > Q_{\rm min} > Q_{\rm thre}$ in order to suppress
(instrumental or physical) backgrounds. Similarly, background rejection may
require a maximum energy cut, $Q < Q_{\rm max} \leq Q_{\rm max,kin}$. In fact,
we will see below that, at least for smallish data samples, such a cut might
even be beneficial for the determination of $m_\chi$. We therefore now give
expressions for the case that only data with $Q_{\rm min} \leq Q \leq Q_{\rm
  max}$ are used in the analysis.

To that end, we introduce generalized moments of $f_1$:
\beqn \label{momnew}
\expv{v^n}(v_1, v_2) \= \int_{v_1}^{v_2} v^n f_1(v) \~ dv
\nonumber \\
\= {\cal N} \alpha^{n+1} \left[ \frac {Q_1^{(n+1)/2} r(Q_1)} {F^2(Q_1)} -
\frac {Q_2^{(n+1)/2} r(Q_2)} {F^2(Q_2)}+ \afrac {n+1} {2} I_n(Q_1,Q_2) 
\right] \, ,
\eeqn
where we have introduced the short--hand notation
\beq \label{r}
r(Q_i) = \left. \frac {dR} {dQ} \right|_{Q=Q_i}\, ,
~~~~~~~~~~~~~~~~~~~~ 
i=1,~2\, ,
\eeq
and 
\beq \label{In}
I_n(Q_1,Q_2) = \int_{Q_1}^{Q_2} Q^{(n-1)/2} \bdRdQoFQ dQ \, .
\eeq
In order to arrive at the final expression in Eq.(\ref{momnew}) we used
Eq.(\ref{eqn2106}) and integrated by parts. The $Q_i$ in Eqs.(\ref{momnew})--(\ref{In})
are related to the original integration limits $v_i$ appearing
on the left--hand side of Eq.(\ref{momnew}) via Eq.(\ref{eqn2104}), i.e.,
\beq  \label{Qi}
Q_i = \frac {v_i^2} {\alpha^2}\, ,
~~~~~~~~~~~~~~~~~~~~ 
i=1,~2\, .
\eeq
Of course, Eq.(\ref{momnew}) reduces to Eq.(\ref{eqn2108}) in the limit $v_1
\rightarrow 0, \, v_2 \rightarrow v_{\rm esc}$; note, however, that
Eq.(\ref{momnew}) is also applicable for the case $n = -1$, where the last
term on the right--hand side vanishes. The same restriction on the WIMP
velocity can also be introduced in the normalization constant ${\cal N}$ of
Eq.(\ref{eqn2107}), in which case $\int_{v_1}^{v_2} f_1(v) \~ dv$ is normalized
to unity. Formally this can be treated using Eqs.(\ref{momnew})--(\ref{Qi}) by
demanding $\expv{v^0}(v_1,v_2) = 1$.

\subsection{Experimental implementation}

In order to directly use our results for $f_1(v)$ and for its moments
$\expv{v^n}$ given in Eqs.(\ref{eqn2106}), (\ref{eqn2108}) and (\ref{momnew}),
one needs a functional form for the recoil spectrum $dR/dQ$. In practice this
results usually from a fit to experimental data. However, data fitting
can re--introduce some model dependence and makes the error analysis
more complicated.  Hence, expressions that allow to reconstruct $f_1(v)$ and
its moments directly from the data have been developed \cite{DMDD}. We
started by considering experimental data described by
\beq \label{eqn2201}
     {\T Q_n-\frac{b_n}{2}}
 \le \Qni
 \le {\T Q_n+\frac{b_n}{2}}\, ,
     ~~~~~~~~~~~~ 
     i
 =   1,~2,~\cdots,~N_n,~
     n
 =   1,~2,~\cdots,~B.
\eeq
Here the total energy range has been divided into $B$ bins with central points
$Q_n$ and widths $b_n$. In each bin, $N_n$ events will be recorded. Note that
we assume that the sample to be analyzed only contains signal events, i.e., is
free of background, and ignore the uncertainty on the measurement of the
recoil energy $Q$. Active background suppression techniques \cite{pdg} should
make the former possible. The energy resolution of most existing detectors is
so good that its error will be negligible compared to the statistical
uncertainty for the foreseeable future.

Since the recoil spectrum $dR/dQ$ is expected to be approximately exponential,
we used the following ansatz for the spectrum in the $n-$th bin \cite{DMDD}:
\beq \label{eqn2202}
        \adRdQ_n
 \equiv \adRdQ_{Q \simeq Q_n}
 \simeq \trn \~ {\rm e}^{k_n (Q-Q_n)}
 \equiv \rn  \~ {\rm e}^{k_n (Q-Q_{s,n})}\, .
\eeq
Here $r_n$ is the standard estimator for $dR/dQ$ at $Q = Q_n$,
\beq \label{eqn2203}
r_n = \frac{N_n}{b_n}\, ,
\eeq
$\trn$ is the value of the recoil spectrum at the point $Q = Q_n$,
\beq \label{eqn2204}
\trn \equiv \adRdQ_{Q = Q_n}
 =      r_n \bfrac{k_n b_n/2}{\sinh (k_n b_n/2)}\, ,
\eeq
and $k_n$ is the logarithmic slope of the recoil spectrum in the $n-$th bin.
It can be computed numerically from the average $Q-$value in the $n-$th bin:
\beq \label{eqn2205}
\bQn = \afrac{b_n}{2} \coth\afrac{k_n b_n}{2}-\frac{1}{k_n}\, ,
\eeq
where
\beq \label{eqn2206}
\bQxn{\lambda} \equiv \frac{1}{N_n} \sumiNn \abrac{\Qni-Q_n}^{\lambda}\, .
\eeq
Finally, $Q_{s,n}$ is the shifted point at which the leading systematic error
due to the ansatz in Eq.(\ref{eqn2202}) is minimal \cite{DMDD},
\beq \label{eqn2207}
Q_{s,n} = Q_n+\frac{1}{k_n} \ln\bfrac{\sinh (k_n b_n/2)}{k_n b_n/2}\, .
\eeq
Note that $Q_{s,n}$ differs from the central point of the $n-$th bin, $Q_n$.

Using Eqs.(\ref{momnew}) and (\ref{r}) with $Q_1 = Q_{\rm min}$ and $Q_2 =
Q_{\rm max}$, the generalized $n-$th moment of the velocity distribution
function can be written as
\beq \label{eqn2208} 
\expv{v^n}(v(Q_{\rm min}), v(Q_{\rm max})) = \alpha^n \bfrac  {2 Q_{\rm
    min}^{(n+1)/2} r(Q_{\rm min}) / F^2(Q_{\rm min}) + (n+1) I_n(Q_{\rm min},
  Q_{\rm max})} {2 Q_{\rm min}^{1/2} r(Q_{\rm min}) / F^2(Q_{\rm min}) +
  I_0(Q_{\rm min}, Q_{\rm max})} \, ,  
\eeq
where $v(Q) = \alpha \sqrt{Q}$. Here we have implicitly assumed that $Q_{\rm
  max}$ is so large that terms $\propto r(Q_{\rm max})$ are negligible. We
will see later that this is not necessarily true for $n \geq 1$, since these
moments receive sizable contributions from large recoil energies \cite{DMDD}.
Nevertheless we will show in Sec.~3 that even in that case, Eq.(\ref{eqn2208})
can still be used for determining $m_\chi$. From the ansatz
Eq.(\ref{eqn2202}), the counting rate at $Q_{\rm min}$ can be expressed as
\beq \label{eqn2209}
r(Q_{\rm min}) = r_1 {\rm e}^{k_1 (Q_{\rm min} - Q_{s,1})}\, . 
\eeq
The integral $I_n(Q_{\rm min}, Q_{\rm max})$ defined in Eq.(\ref{In}) can be
estimated through the sum:
\beq \label{eqn2210} 
I_n(Q_{\rm min}, Q_{\rm max}) = \sum_a \frac{Q_a^{(n-1)/2}}{F^2(Q_a)}\, , 
\eeq
where the sum runs over all events in the data set that satisfy $Q_a \in
[Q_{\rm min}, Q_{\rm max}]$.

Since all $I_n$ are determined from the same data, they are correlated, with
\cite{DMDD}
\beq \label{eqn2211}
{\rm cov}(I_n,I_m) = \sum_a \frac{Q_a^{(n+m-2)/2}}{F^4(Q_a)}\, ,  
\eeq
where the sum again runs over all events with recoil energy between $Q_{\rm
  min}$ and $Q_{\rm max}$. 

On the other hand, the statistical error of $r(Q_{\rm min})$ can be obtained
from Eq.(\ref{eqn2209}) as
\beq \label{eqn2212}
\sigma^2(r(Q_{\rm min})) = r^2(Q_{\rm min}) 
\cbrac{ \frac{\sigma^2(r_1)}{r_1^2} + \bbrac{\frac{1}{k_1} - \afrac{b_1}{2} 
\left( 1 + \coth \afrac{b_1  k_1} {2} \right) }^2 \sigma^2(k_1)}\, .
\eeq
The error on $r_1$ follows directly from its definition in Eq.(\ref{eqn2203}):
\beq \label{eqn2213}
\sigma^2(r_n) = \frac{N_n}{b_n^2}\, .
\eeq
The error on the logarithmic slope $k_1$ can be computed from
Eq.(\ref{eqn2205}):
\beq \label{eqn2214}
\sigma^2(k_n) = k_n^2 \cbrac{1-\bfrac{k_n b_n/2}{\sinh (k_n b_n/2)}^2}^{-1}
\sigma^2\abrac{\bQn} \, ,
\eeq
with
\beq \sigma^2\abrac{\bQn} = \frac{1}{N_n-1} \bbigg{\bQQn-\bQn^2}\, .
\eeq

Finally, the correlation between the errors on $r(Q_{\rm min})$, which is
calculated entirely from the events in the first bin, and on $I_n$ is given by
\cite{DMDD}
\beqn \label{eqn2216}
\conti
   {\rm cov}(r(Q_{\rm min}),I_n)
   \non\\
\= r(Q_{\rm min}) \~ I_{n}(Q_{\rm min}, Q_{\rm min} + b_1)
   \non\\
\conti ~~~~ \times 
   \cleft{\frac{\sigma^2(r_1)}{r_1^2} 
  +\bbrac{\frac{1}{k_1} - \afrac{b_1}{2} \left( 1 + \coth \afrac{b_1 k_1} {2} 
\right)   } }
\non\\
\conti ~~~~~~~~~~~~~~ \times 
\cright{\bbrac{\frac{I_{n+2}(Q_{\rm min}, Q_{\rm min}+b_1)}{I_{n}(Q_{\rm min},
    Q_{\rm min} + b_1)} - Q_1 + \frac{1}{k_1} - \afrac{b_1}{2} \coth \afrac{b_1
    k_1} {2}} \sigma^2(k_1)}\, ;
\eeqn
note that the integrals $I_i$ in Eq.(\ref{eqn2216}) only extend over the first
bin, which ends at $Q = Q_{\rm min} + b_1$.

\section{Determining the WIMP mass}

We are now ready to describe methods to extract the WIMP mass $m_\chi$ from
direct detection data. Recall that $m_\chi$ is an {\em input} in the
determination of $f_1$ and its moments as described in the previous
section. In particular, $m_\chi$ appears in the factor $\alpha^n$ on the
right--hand side of Eq.(\ref{eqn2208}) describing the experimental estimate
of the moments of $f_1$. A truly model--independent determination of $m_\chi$
from these data will therefore only be possible by requiring that two (or
more) experiments, using different target nuclei, lead to the same result for
$f_1$. 

Here we focus on the moments of the distribution function, rather than the
function itself, since non--trivial information about the former can already
be obtained with ${\cal O}(20)$ events \cite{DMDD}. We will also show how
$m_\chi$ can be estimated from the knowledge of the integral $I_0$ appearing
in the normalization of $f_1$, if the ratio of WIMP scattering cross sections
on protons and neutrons is known. For greater clarity, we will first discuss
the idealized situation where all signal events, irrespective of their recoil
energy $Q_a$, are included. In the second subsection we will introduce upper
and lower limits on $Q_a$. A third subsection is devoted to a discussion how
the different estimators for the WIMP mass can be combined using a $\chi^2$
fit.

\subsection{Without cuts on the recoil energy}

If no cuts on the recoil energy need to be applied, we can use the original
moments, or equivalently, the generalized moments with $v_1 \rightarrow 0$ and
$v_2 \rightarrow v_{\rm esc}$; recall that the latter is the same as allowing
$v_2 \rightarrow \infty$, since by assumption $f_1(v) = 0$ for $v > v_{\rm
  esc}$. Eq.(\ref{eqn2208}) then simplifies to
\beq \label{eqn3101}
\expv{v^n} = \alpha^n (n+1) \afrac{I_n}{I_0}\, ,
\eeq
where $I_n$ and $I_0$ can be estimated as in Eq.(\ref{eqn2210}) with $Q_{\rm
  min} \rightarrow 0$ and $Q_{\rm max} \rightarrow \infty$ (or, equivalently,
$Q_{\rm max} \rightarrow Q_{\rm max,kin}$).

Suppose $X$ and $Y$ are two target nuclei. We denote their masses by $m_X$,
$m_Y$, and their form factors as $F_X(Q)$, $F_Y(Q)$. Similarly, we define
$\alpha_{X,Y}$ as in Eq.(\ref{eqn2105}), with $m_N \rightarrow m_{X,Y}$.
Eq.(\ref{eqn3101}) then implies
\beq \label{eqn3102}
\alpha_X^n \afrac{\InX}{\IzX} = \alpha_Y^n \afrac{\InY}{\IzY}\, .
\eeq
Note that the form factors in the estimates of $\InX$ and $\InY$ using
Eq.(\ref{eqn2210}) are different. Eq.(\ref{eqn3102}) can be solved for
$m_\chi$ using Eqs.(\ref{eqn2105}) and (\ref{eqn2103}):
\beq \label{eqn3103}
\mchi = \frac {\sqrt{\mX \mY} - \mX \calRn} {\calRn - \sqrt{\mX/\mY}}\, , 
\eeq
where we have defined
\beq \label{eqn3104} 
\calRn \equiv \frac{\alpha_Y}{\alpha_X} =
\abrac{\frac{\InX}{\IzX} \cdot \frac{\IzY}{\InY}}^{1/n}\, ,
~~~~~~~~~~~~~~~~~~~~   n \ne    0,~-1.
\eeq

Using standard Gaussian error propagation, the statistical error on $m_\chi$
estimated from Eq.(\ref{eqn3103}) is
\beqn \label{eqn3105}
 \left.\sigma(\mchi)\right|_{\Expv{v^n}} 
\=  \frac{\sqrt{\mX/\mY} \vbrac{\mX-\mY}} {\abrac{\calRn -
    \sqrt{\mX/\mY}}^2} \cdot \sigma(\calRn)
\non\\
 \= \frac{\calRn \sqrt{\mX/\mY} \vbrac{\mX-\mY}} {\abrac{\calRn -
     \sqrt{\mX/\mY}}^2} 
        \non\\
 \conti ~~~~~~ \times 
 \frac{1}{|n|} \bbrac{ \frac{\sigma^2(\InX)}{\InX^2}
 + \frac{\sigma^2(\IzX)}{\IzX^2}
 - \frac{2 {\rm cov}(\IzX,\InX)}{\IzX \InX}
 + (X \lto Y)}^{1/2}\, ,
\eeqn
where $\sigma^2(\InX) = {\rm cov}(\InX,\InX)$ and ${\rm cov}(\IzX,\InX)$ and
so on can be estimated from Eq.(\ref{eqn2211}).

A second method for determining $m_\chi$ starts directly from
Eq.(\ref{eqn2101}), plus an assumption about the relative strength for WIMP
scattering on protons p and neutrons n. The simplest such assumption is that the
scattering cross section is the same for both nucleons. This is in fact an
excellent approximation for the spin--independent contribution to the cross
section of supersymmetric neutralinos \cite{susydm}, and for all WIMPs which
interact primarily through Higgs exchange. Writing the ``pointlike'' cross
section $\sigma_0$ of Eq.(\ref{eqn2102}) as
\beq \label{sigma0}
\sigma_0 = \Afrac{4}{\pi} m_{\rm r,N}^2 A^2 |f_{\rm p}|^2\, ,
\eeq
where $f_{\rm p}$ is the effective $\chi \chi {\rm p p}$ 4--point coupling, $m_{\rm r,N}$ is the
reduced mass defined in Eq.(\ref{eqn2103}) and $A$ is the number of nucleons in the nucleus, we have from
Eqs.(\ref{eqn2101}), (\ref{eqn2102}) and the first Eq.(\ref{momnew}):
\beq \label{rQmin}
r(Q_{\rm min}) = \frac {\rho_0} {2 m_\chi} \Afrac{4}{\pi} A^2 |f_{\rm p}|^2 F^2(Q_{\rm min}) 
\expv{v^{-1}}(v(Q_{\rm min}), v_{\rm esc}) \, .
\eeq
Using the second Eq.(\ref{momnew}) we see that the counting rate at $Q_{\rm
  min}$ in fact drops out, and we are left with
\beq \label{one}
1 = \frac { 4 \sqrt{2}}{\pi} \afrac{{\cal E} \rho_0 A^2 |f_{\rm p}|^2}{I_0}
\afrac{m_{\rm r,N}}{m_\chi \sqrt{\mN}}\, ,
\eeq
where we have assumed $Q_{\rm min} = 0$. Recall that the rate $dR/dQ$ is
defined as rate per unit mass and observation time interval, i.e., we need to
divide the actual event rate by the exposure ${\cal E} = M \tau$, where $M$ is
the (fiducial) mass of the detector and $\tau$ the observation time. In our
previous discussion this factor always dropped out in the end, due to the
appearance of the normalization ${\cal N}$. This is true also for the
right--hand side of Eq.(\ref{rQmin}), but not for the $1/{\cal E}$ factor on
the left--hand side of this equation; hence a factor ${\cal E}$ appears in the
numerator of Eq.(\ref{one}). Note that ${\cal E}$ is dimensionless in natural
units. On the other hand, the unknown factor $\rho_0 |f_{\rm p}|^2$ appearing in
Eq.(\ref{one}) will cancel out when we use this identity for two different
targets $X$ and $Y$, leading to the final result
\beq \label{msigma}
m_\chi =
 \frac{\abrac{\mX/\mY}^{5/2} \mY-\mX \calR_{\sigma}}{\calR_{\sigma}-\abrac{\mX/\mY}^{5/2}}\, .
\eeq
Here we have assumed $m_{X,Y} \propto A_{X,Y}$, and introduced the quantity
\beq \label{Rsigma}
{\cal R}_\sigma = \frac{{\cal E}_Y}{{\cal E}_X} \afrac{I_{0,X}}{I_{0,Y}}\, .
\eeq

Recall that, even though the derivation started from the expression for the
counting rate at $Q_{\rm min}$, this rate actually dropped out when going from
Eq.(\ref{rQmin}) to Eq.(\ref{one}). The final expression only depends on the
quantity $R_{\sigma}$, which is estimated from {\em all} the events in both
samples using Eq.(\ref{eqn2210}). The error on $m_\chi$ computed from
Eq.(\ref{msigma}) is therefore comparable to that for $m_\chi$ derived from a
moment of $f_1$. Still keeping $Q_{\rm min} = 0$, we have
\beq \label{sigsig}
\left. \sigma(m_\chi) \right|_\sigma
 = \frac{\calR_{\sigma} \abrac{\mX/\mY}^{5/2} \vbrac{\mX-\mY}}
        {\bbrac{\calR_{\sigma}-\abrac{\mX/\mY}^{5/2}}^2}
   \bbrac{\frac{\sigma^2\abrac{\IzX}}{\IzX^2}+\frac{\sigma^2\abrac{\IzY}}{\IzY^2}}^{1/2}\, .
\eeq
Ultimately the ${\cal R}_n,\, {\cal R}_\sigma$ and the errors in
Eqs.(\ref{eqn3105}) and (\ref{sigsig}) should be estimated from the data
directly. In the meantime it is instructive to note that the final expressions
for the statistical errors of our estimators for $m_\chi$ decompose into two
factors. The expectation value of the first factor does not depend on $f_1$;
it can be computed entirely from the masses of the involved particles:
\beqn \label{kappa}
\frac { {\cal R}_n \sqrt{m_X / m_Y} \left| m_X - m_Y \right|}
{ \left( {\cal R}_n - \sqrt{ m_X / m_Y } \right)^2 } 
\= \frac{\calR_{\sigma} \abrac{\mX/\mY}^{5/2} \vbrac{\mX-\mY}}
        {\bbrac{\calR_{\sigma}-\abrac{\mX/\mY}^{5/2}}^2}
   \non\\
\= \frac { \left( m_\chi + m_X \right) \left( m_\chi + m_Y \right) } 
{ \left| m_X - m_Y \right| }\non\\
 \eqnequiv \kappa \, .
\eeqn
Here we have made use of the identities ${\cal R}_n = \alpha_Y /\alpha_X$ in Eq.(\ref{eqn3104})
and
\beq
{\cal R}_\sigma = \left( \frac {m_X} {m_Y} \right)^{5/2} 
\afrac { m_\chi + m_Y } { m_\chi + m_X }\, ,
\eeq
which hold for the expectation values of these quantities as can be seen from
Eqs.(\ref{eqn3103}) and (\ref{msigma}).  Remarkably, the expectation values of
the factors in front of the expressions in square brackets are in fact the
same in Eqs.(\ref{eqn3105}) and (\ref{sigsig}), apart from the factor $1/|n|$
appearing in the former equation.  On the other hand, the expressions inside
these square parentheses do depend on $f_1(v)$, as well as on the involved
masses and form factors.

It is nevertheless instructive to study the behavior of the factor $\kappa$
for different target nuclei $X$ and $Y$. This largely determines what choice
of targets is optimal, i.e., minimizes the statistical errors of our estimators
of $m_\chi$. It is easy to see that the ratio $\kappa/m_\chi$, which will
appear in the relative error $\sigma(m_\chi) / m_\chi$, only depends on the
dimensionless ratios $m_X / m_\chi$ and $m_Y / m_\chi$. In
Fig.~1 
we therefore show contours of $\kappa/m_\chi$ in the
plane spanned by these two ratios, using the ordering $m_Y > m_X$. We note
first of all that $\kappa$ diverges as $m_Y \rightarrow m_X$. This is clear
from the fact that the denominator in the final expression in Eq.(\ref{kappa})
vanishes in this limit, while the numerator is finite. Physically this simply
means that performing two scattering experiments with the same target will not
allow one to determine $m_\chi$.

\begin{figure}[t!] \label{fig:contour}
\begin{center}
\vspace*{-1cm}
\rotatebox{-90}{\includegraphics[width=13.5cm]{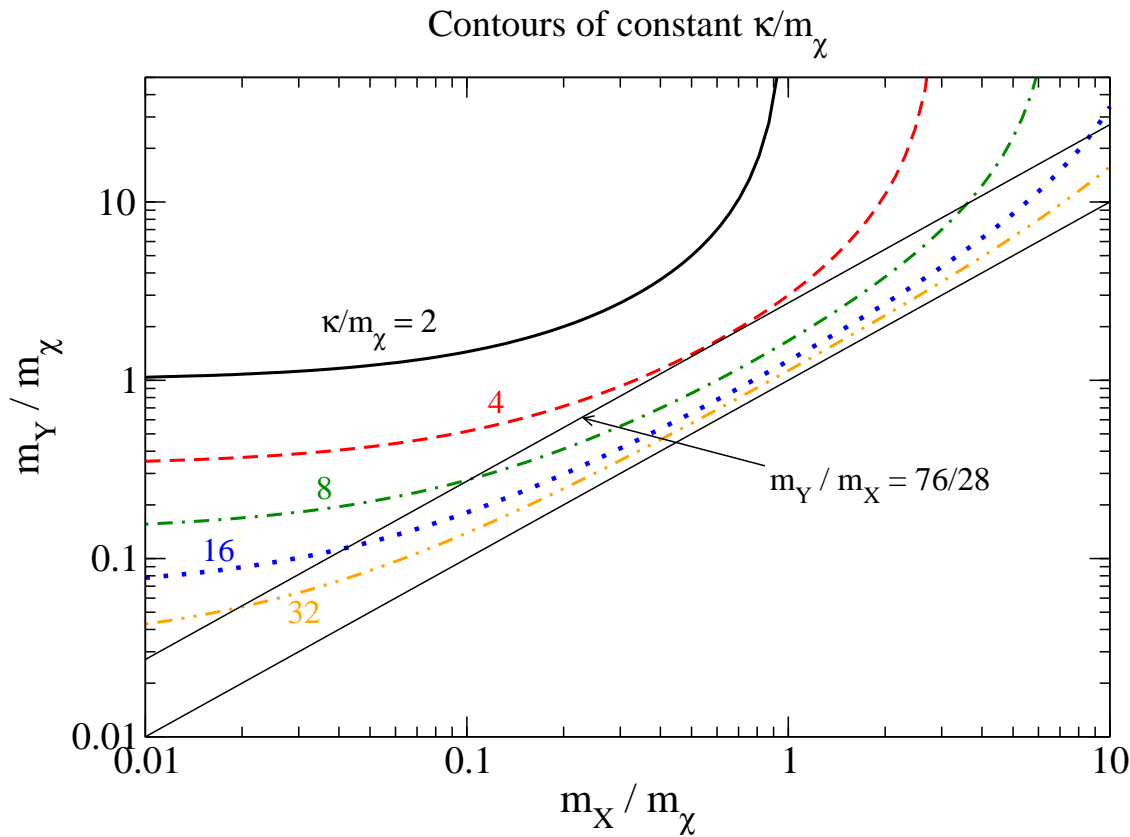}}
\end{center}
\vspace*{-1cm}
\caption{The five thick lines show contours of constant $\kappa/m_\chi$ in the
  plane spanned by $m_X / m_\chi$ and $m_Y / m_\chi$, where we have taken $m_Y
  > m_X$ without loss of generality. Here $\kappa$, has been defined in
  Eq.(\ref{kappa}); recall that the final relative error on $m_\chi$ is
  directly proportional to $\kappa / m_\chi$. The lower thin line indicates
  the end of the physical region, $m_Y = m_X$, whereas the upper thin line
  shows $m_Y = (76/28) m_X$, corresponding to Silicon and Germanium targets.}
\end{figure}

Slightly less trivially, we also see that $\kappa/m_\chi$, and hence the
relative error on $m_\chi$, becomes very large if both target nuclei are
either much heavier or both much lighter than the WIMP. This can be understood
from the fact that $m_\chi$ only enters via the reduced mass defined in
Eq.(\ref{eqn2103}).\footnote{Eq.(\ref{rQmin}) seems to have additional
  $m_\chi$ dependence. However, this comes from the factor $n_\chi = \rho_0 /
  m_\chi$, which drops out when the ratio of two targets is considered.} If
$m_\chi \gg \mN, \ m_{\rm r,N} \rightarrow \mN$ becomes completely independent of
the WIMP mass. In the opposite extreme, $m_\chi \ll \mN,\ m_{\rm r,N} \rightarrow
m_\chi$ becomes independent of the mass of the target nucleus, i.e., one is
effectively back in the situation where both experiments are performed with
the same target.

These considerations favor chosing the two targets to be as different as
possible. However, there are limits to this. On the one hand, taking a very
light target nucleus will lead to a low event rate for this experiment, and
hence very large statistical errors. Since the errors of both experiments are
added in quadrature inside the expressions in square parenthesis in
Eqs.(\ref{eqn3105}) and (\ref{sigsig}), this would lead to a large overall
error on $m_\chi$. In fact, if the total event number is held fixed, the final
error on $m_\chi$ will be minimal if both experiments contain approximately
the same number of events.

On the other hand, taking a very heavy target nucleus also leads to problems.
Heavy nuclei are large, which means they have quite soft form factors. For
example, the Woods--Saxon prediction \cite{FQb} for the form factor for
${}^{136}$Xe, which is the target in some existing experiments \cite{pdg}, has
a zero at $Q \simeq 95$ keV. For our default choice $v_{\rm esc} = 700$ km/s,
this is below $Q_{\rm max,kin}$ of Eq.(\ref{qmaxkin}) for all $m_\chi > 45$
GeV. This is a serious problem, since the form factor, and hence the event
rate, remain (very) small even beyond this zero. Experiments with ${}^{136}$Xe
therefore effectively impose a cut $Q_{\rm max} \leq 95$ GeV.\footnote{Note
  also that while the integrals $I_n$ remain finite where the form factor
  vanishes, the estimates for the errors will diverge there, due to the factor
  $1/F^4$ appearing in Eq.(\ref{eqn2211}).}

From these considerations it seems that chosing $X =$ Si, $Y =$ Ge might be a
good compromise. This corresponds to points on the upper, thin solid straight
line in Fig.~1. 
We see that in this case $\kappa / m_\chi \geq
4$ for all values of $m_\chi$. Not surprisingly, $\kappa/m_\chi$ reaches its
minimum when $m_\chi$ lies in between the masses of the two target nuclei, i.e
for the case at hand, for $m_\chi \simeq 50$ GeV. Note also that
$\kappa/m_\chi$ is unfortunately always well above unity. One will thus have
to get fairly accurate estimates of the relevant integrals $I_i$ if one wishes
to determine $m_\chi$ to better than a factor of two.

Before discussing the statistical uncertainty in more detail, we proceed to
include non--trivial upper and lower cuts on the recoil energy in our analysis.

\subsection{Incorporating cuts on the recoil energy}

As noted earlier, real experiments will probably have to impose both lower and
upper limits on the recoil energy, partly for instrumental reasons, and partly
to suppress backgrounds. This led us to introduce generalized moments of $f_1$
in Eq.(\ref{momnew}). These moments determined by two different detectors will
still be the same, if either the integration limits are identical, \mbox{$v_{i,X} =
v_{i,Y}$}, $i = 1,~2$, or the cuts are so loose that the contribution from WIMPs
with $v < v_1$ or $v > v_2$ is negligible; a combination of these two
possibilities can also occur, e.g., small but not necessarily equal values of
$v_{1,X}$ and $v_{1,Y}$, and $v_{2,X} = v_{2,Y}$ significantly below $v_{\rm
  esc}$. 

The crucial observation that forms the basis for our analysis of $m_\chi$ as
estimated from the generalized moments is that in fact {\em all three} terms
in the final expression in Eq.(\ref{momnew}) will agree {\em separately}
between two targets, as long as both integration limits coincide. This can be
shown by replacing the $r(Q_i)$ in Eq.(\ref{momnew}) via Eq.(\ref{eqn2101})
and using Eq.(\ref{Qi}). In principle we could therefore simply equate the
last (integral) terms between the two targets. This would lead to expressions
very similar to those derived in the previous subsection, the only difference
being that all integrals or sums over recoil energies would run over limited
ranges. 

The problem with this approach is that the velocities $v_i$ appearing in
Eq.(\ref{momnew}) are not directly observable. The recoil energies are.
However, the values of $Q_{\rm min}$ and $Q_{\rm max}$ would need to be {\em
  different} for the two targets, if they are to correspond to the same values
of $v_1$ and $v_2$. Worse, Eq.(\ref{Qi}) shows that the ratios $Q_{{\rm
    min},X} / Q_{{\rm min},Y}$ and $Q_{{\rm max},X} / Q_{{\rm max},Y}$ that
one has to chose in order to achieve $v_{i,X} = v_{i,Y}$ depend on
$m_\chi$. We are thus faced with the situation that, in the presence of
significant cuts on the recoil energy, the quantity we wish to determine is
needed as an input at an early stage of the analysis!

This statement is, strictly speaking, true; we see no way around this, if
significant cuts on the recoil energy are indeed needed. However, we can
alleviate the situation. To begin with, we noticed that the systematic error
one introduces by simply setting $Q_{{\rm min},X} = Q_{{\rm min},Y}$ is
reduced greatly if one keeps the sum of the first and third terms in
Eq.(\ref{momnew}). Moreover, we will show that, to some extent at least, one
can do the matching of two $Q_{\rm max}$ values for the two targets directly
from the data. In principle we could reduce the systematic error associated
with the choice of $Q_{\rm max}$ even more by also including the second term
in Eq.(\ref{momnew}). However, with limited statistics the error on $r(Q_{\rm
  max})$ will always be very large, so that keeping this term will not help
significantly.

Our determination of $m_\chi$ from generalized moments of $f_1$ is thus based
on equating the sum of the first and third terms in Eq.(\ref{momnew}) for two
different targets. The resulting expression for $m_\chi$ can still be cast in
the form of Eq.(\ref{eqn3103}), but ${\cal R}_n$ is now given by
\beq \label{eqn3201}
\calRn(Q_{\rm min}) = \bfrac{2 Q_{{\rm min},X}^{(n+1)/2} 
r_X(Q_{{\rm min},X})/F_X^2(Q_{{\rm min},X}) + (n+1) \InX}
{2 Q_{{\rm min},X}^{1/2} r_X(Q_{{\rm min},X})/ F_X^2(Q_{{\rm min},X}) + \IzX}^{1/n}
 (X \lto Y)^{-1}\, .
\eeq
Here $n \ne 0$
and $r_{(X,Y)}(Q_{{\rm min},(X,Y)})$ refer to the counting rate for detectors
$X$ and $Y$ at the respective lowest recoil energies included in the analysis.
They can be estimated from Eq.(\ref{eqn2209}); of course, the values of $r_1$,
$k_1$, $Q_{s,1}$, and $b_1$ will in general differ for the two targets. Note
that the integrals $I_n, \, I_0$ are now to be evaluated with $Q_{\rm min}$ as
lower and $Q_{\rm max}$ as upper limits, as in Eq.(\ref{eqn2210}).

\begin{figure}[b!] \label{fig:mrec_qmin}
\begin{center}
\vspace*{-1cm}
\rotatebox{-90}{\includegraphics[width=13.5cm]{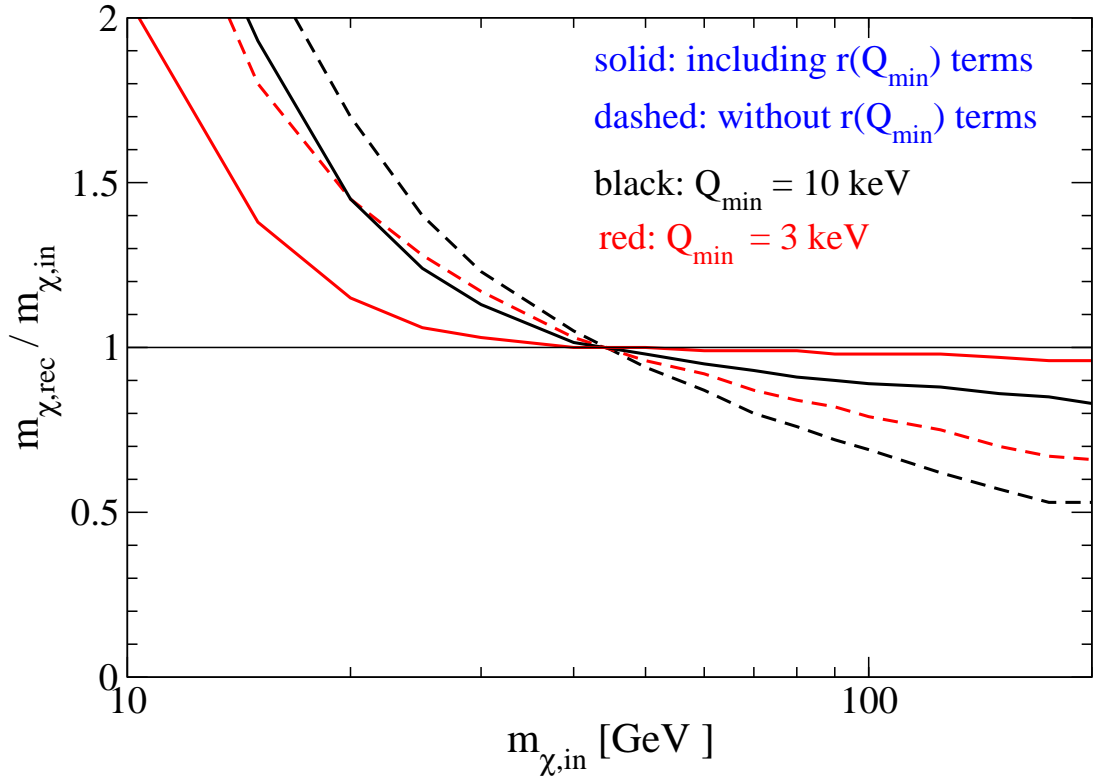}}
\end{center}
\vspace*{-1cm}
\caption{Ratio of reconstructed to input WIMP mass for imperfect $Q_{\rm min}$
  matching, for two different values of $Q_{\rm min}$. We have taken Si and Ge
  targets and assumed infinite statistics and (effectively) no upper cut on
  the recoil energy. The solid (dashed) curves have been computed from
  Eqs.(\ref{eqn3103}) and (\ref{eqn3201}) with $n=1$ including (neglecting)
  all terms $\propto r(Q_{\rm min})$. See the text for further details.}
\end{figure}

Before discussing the statistical uncertainty of this estimate of the WIMP
mass, we wish to demonstrate that keeping the terms $\propto r(Q_{\rm min})$
in Eq.(\ref{eqn3201}) indeed reduces the systematic error.  This is
demonstrated in Fig.~2, 
which shows the ratio of the
reconstructed to input WIMP mass for experiments with ``infinite'' statistics.
Here we have chosen $X = {}^{28}$Si and $Y = {}^{76}$Ge, and we have set
$Q_{\rm max} = \infty$ in order to isolate the systematic effect due to
imperfect matching of the $Q_{\rm min}$ values in this figure.\footnote{More
  exactly, for $m_\chi > 100$ GeV we have set $Q_{\rm max, \, Ge} = 250$ keV,
  in order to avoid complications due to the zero of the form factor of
  ${}^{76}$Ge which occurs at $Q \simeq 270$ keV. We have then matched $Q_{\rm
    max, \, Si}$, so that the only systematic deviation of the reconstructed
  WIMP mass is indeed due to imperfect $Q_{\rm min}$ matching.} Here the black
(red) curves have been obtained with $Q_{\rm min, \, Ge} = Q_{\rm min, \, Si}
= 10 \ (3)$ keV. The solid lines include the terms $\propto r(Q_{\rm min})$,
while the dashed ones do not. Clearly including these terms is beneficial: for
$m_\chi > 20$ GeV, the systematic shift with these terms included and $Q_{\rm
  min} = 10$ keV is smaller than that without those terms with the much
smaller $Q_{\rm min} = 3$ keV. We also see that the systematic effect becomes
large at small $m_\chi$. This is not surprising, since a small $m_\chi$
implies a small $Q_{\rm max,kin}$, so that the lower cut on $Q$ removes a
correspondingly larger fraction of the total spectrum. The systematic error
vanishes for $m_\chi = \sqrt{m_X m_Y} = 43$ GeV, since for this value of
$m_\chi$ we have $\alpha_X = \alpha_Y$, i.e., $v_{1,X} = v_{1,Y}$ if $Q_{{\rm
    min},X} = Q_{{\rm min}, Y}$. For larger $m_\chi$ the systematic effect
changes sign, i.e., one now underestimates the true value of $m_\chi$.
However, if the terms $\propto r(Q_{\rm min})$ are included, the shift remains
relatively small even for $Q_{\rm min} = 10$ keV. If values $Q_{\rm min} \lsim
3$ keV can be achieved, which should be possible \cite{low_th}, the systematic
effect due to imperfect $Q_{\rm min}$ matching will be a concern only for very
light WIMPs, $m_\chi < 20$ GeV.

Using Eq.(\ref{eqn3201}) in Eq.(\ref{eqn3103}) also leads to a lengthier
expression for the statistical error on our estimate of $m_\chi$. The first
equation in (\ref{eqn3105}) still holds, but the $r(Q_{\rm min})$ terms
give additional contributions to the final expression:
\beqn \label{sigmom}
\left. \sigma(m_\chi)\right|_{\Expv{v^n}} \= 
\frac{\sqrt{\mX/\mY} \vbrac{\mX-\mY}} {\abrac{\calRn -
     \sqrt{\mX/\mY}}^2}
\non\\
\conti ~~~~~~~~~~~~ \times 
 \left[ \sum_{i,j=1}^3 \afrac {\partial {\cal R}_n}
{\partial c_{i,X}} \afrac {\partial {\cal R}_n} {\partial c_{j,X} } {\rm
  cov}(c_{i,X}c_{j,X}) + (X \lto Y) \right]^{1/2}\, .
\eeqn
Here we have introduced a short--hand notation for the six quantities on which
the estimate of $m_\chi$ depends:
\beq \label{ci}
c_{1,X} = I_{n,X}\, ; ~~~~~~~~~~~~ 
c_{2,X} = I_{0,X}\, ; ~~~~~~~~~~~~ 
c_{3,X} = r_X(Q_{{\rm min},X})\, ,
\eeq
and similarly for the $c_{i,Y}$. Estimators for ${\rm cov}(c_i, c_j)$ have
been given in Eqs.(\ref{eqn2211}) and (\ref{eqn2216}). Explicit expressions
for the derivatives of ${\cal R}_n$ with respect to these six quantities are
collected in the Appendix. Note that a factor ${\cal R}_n$ can be factored
out of all these derivatives. With this factor moved out of the square
brackets, the prefactor in Eq.(\ref{sigmom}) is identical to that in
Eq.(\ref{eqn3105}), i.e., its expectation value will again be given by $\kappa$
of Eq.(\ref{kappa}). Of course, Eq.(\ref{sigmom}) reduces to
Eq.(\ref{eqn3105}) in the limit $Q_{{\rm min},X}\, , Q_{{\rm min},Y}
\rightarrow 0$. Note finally that Eq.(\ref{sigmom}) also holds for $n = -1$,
if the derivatives with respect to $c_{1,(X,Y)}$ are neglected, since in this
case $\InX$ and $\InY$ do not contribute to ${\cal R}_n$ given in
Eq.(\ref{eqn3201}).

Finite lower energy cuts $Q_{{\rm min},(X,Y)}$ can also easily be incorporated
in the quantity ${\cal R}_\sigma$ appearing in Eq.(\ref{msigma}):
\beq \label{Rsigma1}
R_\sigma = \frac {{\cal E}_Y} {{\cal E}_X}
\bfrac{2Q_{{\rm min},X}^{1/2} r_X(Q_{{\rm min},X}) / F_X^2(Q_{{\rm min},X}) + \IzX}
{2 Q_{{\rm min},Y}^{1/2} r_Y(Q_{{\rm min},Y}) / F_Y^2(Q_{{\rm min},Y}) + \IzY} 
\,.
\eeq
Correspondingly, Eq.(\ref{sigsig}) changes to
\beqn \label{sigsig1}
\left. \sigma(m_\chi) \right|_\sigma \=
\frac{\abrac{\mX/\mY}^{5/2} \vbrac{\mX-\mY}}
     {\bbrac{\calR_{\sigma}-\abrac{\mX/\mY}^{5/2}}^2}
\non\\
\conti ~~~~~~~~~~~~ \times 
 \left[ \sum_{i,j=2}^3 \afrac
  {\partial {\cal R}_\sigma} {\partial c_{i,X}}
  \afrac {\partial {\cal R}_\sigma}  {\partial c_{j,X} } 
{\rm cov}(c_{i,X}c_{j,X}) + (X \lto Y) \right]^{1/2}\, ,
\eeqn
where we have again used the short--hand notation of Eq.(\ref{ci}); note that
$c_{1(X,Y)}$ does not appear here. Expressions for the derivatives of ${\cal
  R}_\sigma$ are also given in the Appendix.

\subsection{Combined fit}

In the next step we wish to combine our estimators (\ref{eqn3103}) for
different $n$ with each other, and with the estimator (\ref{msigma}). This
could be done via an overall covariance matrix describing the errors of these
estimators and their correlations. The diagonal entries of this covariance
matrix are given by Eqs.(\ref{sigmom}) and (\ref{sigsig1}); the off--diagonal
entries can be computed analogously. This would yield the overall best--fit
value of $m_\chi$ as well as its Gaussian error.

Here we pursue a slightly different procedure, based on a $\chi^2$ fit. This
will yield the same best--fit value of $m_\chi$, which we denote by $m_{\chi,
  {\rm rec}}$, but it has two advantages. First, $\chi^2(m_{\chi,{\rm rec}})$
can be used as a measure of the quality of the fit, which in turn can be used
to match the $Q_{\rm max}$ values of the two experiments at least
approximately. Secondly, it allows to determine asymmetric error intervals.
Fig.~1 
implies that the errors should indeed be asymmetric.
For example, if the true $m_\chi$ is (much) larger than the masses of both
target nuclei, the experimental upper bound one can derive will be quite large
or even infinite, but one should still get a meaningful lower bound; the
opposite is true if $m_\chi$ lies well below the mass of both target nuclei.

We begin by defining fit functions
\cheqna
\beq \label{fia}
f_{i,X} = \alpha_X^i \bfrac  {2 Q_{{\rm min},X}^{(i+1)/2} 
r_X(Q_{{\rm min},X}) /
  F_X^2(Q_{{\rm min},X}) + (i+1) I_{i,X}} {2
  Q_{{\rm min},X}^{1/2} r_X(Q_{{\rm min},X}) / F_X^2(Q_{{\rm min},X}) + 
  I_{0,X}}
 \afrac {300 \ {\rm km}} {\rm s}^{-i} \, ,
\eeq
 for $i=-1,~1,~2,~\dots,~n_{\rm max}$;
 and
\cheqnb
\beq \label{fib}
f_{n_{\rm max}+1,X} = {\cal E}_X \bfrac {A_X^2} 
{2 Q_{{\rm min},X}^{1/2} r_X(Q_{{\rm min},X}) / F_X^2(Q_{{\rm min},X}) + \IzX}
\afrac{\sqrt{\mX}}{\mchi+\mX}
\, ;
\eeq
\cheqn
we analogously define $n_{\rm max} + 2$ functions $f_{i,Y}$.  Here $n_{\rm
  max}$ determines the highest (generalized) moment of $f_1$ that is included
in the fit. The $f_i$ are normalized such that they are dimensionless and very
roughly of order unity; this alleviates numerical problems associated with the
inversion of their covariance matrix. The first $n_ {\rm max}+1$ functions
$f_i$ are basically our estimators (\ref{eqn2208}) of the generalized moments
defined in Eq.(\ref{momnew}) with the term $\propto r(Q_{\rm max})$ omitted;
as discussed earlier, the error on this quantity is likely to be so large that
including this term will not be helpful. The last function is essentially the
ratio appearing in Eq.(\ref{one}). It is important to note that $m_\chi$ in
Eqs.(\ref{fia}) and (\ref{fib}) is a fit parameter, not the true (input) value of the WIMP
mass. Recall also that our estimator (\ref{eqn2210}) for the integrals $I_n$
appearing in Eqs.(\ref{fia}) and (\ref{fib}) is independent of $m_\chi$.
Hence the first $n_{\rm max} + 1$ fit functions depend on $m_\chi$
only through the overall factor
$\alpha^i$.

The $f_i$ allow us to introduce a $\chi^2$ function:
\beq \label{chisq}
\chi^2 = \sum_{i,j} \left( f_{i,X} - f_{i,Y} \right) {\cal C}^{-1}_{ij}
\left( f_{j,X} - f_{j,Y} \right) \, .
\eeq
Here ${\cal C}$ is the total covariance matrix. Since the $X$ and $Y$
quantities are statistically completely independent, ${\cal C}$ can be written
as a sum of two terms:
\beq \label{C}
{\cal C}_{ij} = {\rm cov}\left( f_{i,X}, f_{j,X} \right) 
+ {\rm cov}\left( f_{i,Y}, f_{j,Y} \right)\,.
\eeq
The entries of this matrix involving only the moments of the WIMP velocity
distribution can be read off Eq.(82) of Ref.~\cite{DMDD}, with an obvious
modification due to the normalization factor in Eq.(\ref{fia}). Since
the last $f_i$ can be computed from the same basic quantities, i.e., the
counting rates at $Q_{\rm min}$ and the integrals $I_0$, the entries of the
covariance matrix involving this last fit function can also be computed
straightforwardly, using Eqs.(\ref{eqn2211})--(\ref{eqn2216}). Of course,
Eq.(\ref{chisq}) can also be used to compute asymmetric error intervals from a
single moment, by restricting the sum to a single term.

\section{Numerical results}

We are now ready to present some numerical results for the reconstructed WIMP mass.
These results are based on Monte Carlo simulations of direct detection
experiments. We assume that the scattering cross section is dominated by
spin--independent interactions, and use the Woods--Saxon form for the elastic
form factors $F(Q)$ \cite{FQb}. We describe the WIMP velocity distribution by
a sum of a shifted Gaussian contribution \cite{susydm} and a ``late infall''
component \cite{sikivie}; as in Ref.~\cite{DMDD}, for simplicity we model the
latter as a $\delta-$function, keeping the normalization $N_{\rm l.i.}$ of
this component as free parameter:
\beq \label{f1}
f_1(v) = \frac {\left( 1 - N_{\rm l.i.} \right)} {\sqrt{\pi}} \afrac {v} {v_e
  v_0} \left[ {\rm e}^{-(v-v_e)^2/v_0^2} - {\rm e}^{-(v+v_e)^2/v_0^2} \right]
+ N_{\rm l.i.} \delta(v - v_{\rm esc}) \,.
\eeq
We take $v_0 = 220$ km/s, $v_e = 1.05 \~ v_0$\footnote{Strictly speaking, $v_e$
  should oscillate around $1.05 \~ v_0$ with a period of one year \cite{susydm};
  we ignore this time dependence here.}, and $v_{\rm esc} = 700$ km/s. 

\begin{figure}[t!] \label{fig:nocut}
\begin{center}
\vspace*{-.25cm}
\rotatebox{-90}{\includegraphics[width=11.5cm]{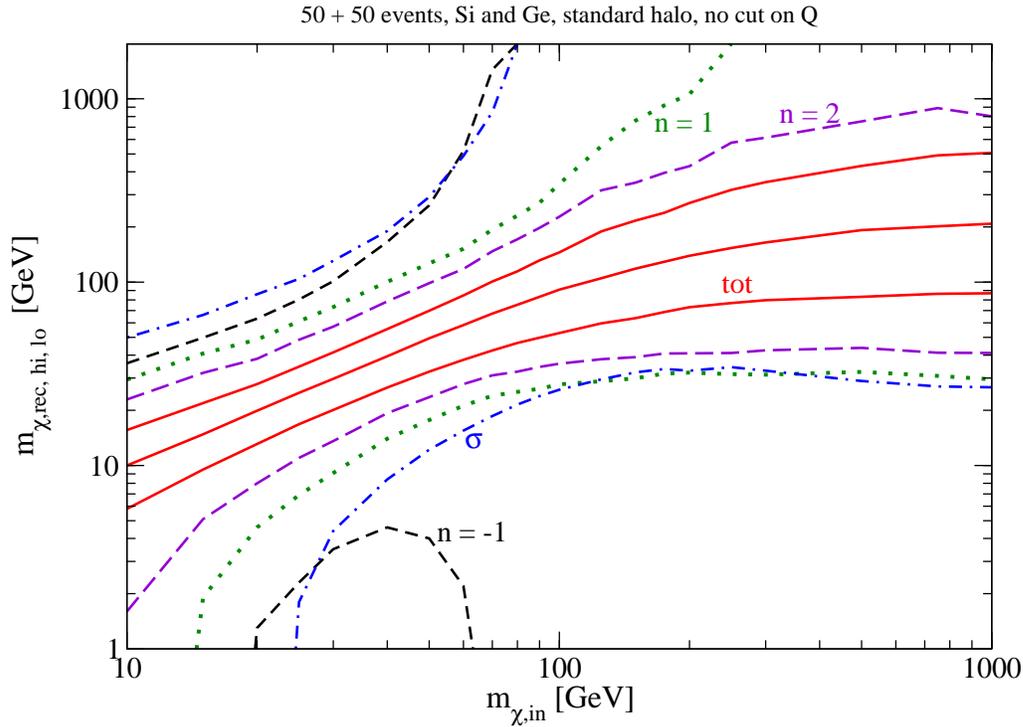}}
\end{center}
\vspace*{-1cm}
\caption{Median values of ``$1\, \sigma$'' upper and lower bounds on the
  reconstructed WIMP mass in $2 \times 5,000$ simulated experiments with Si
  and Ge targets, as a function of the true value of $m_\chi$. We generated on
  average 50 events per experiment. The short--dashed (black), dotted (green)
  and long--dashed (violet) curves show the upper and lower bounds on $m_\chi$
  as determined from moments with $n=-1, \, 1$ and 2, respectively; the
  dash-dotted (blue) curves labeled $\sigma$ show the corresponding limits derived
  from our assumption of equal cross sections for scattering on protons and
  neutrons. Finally, the solid (red) curves show the upper and lower bound as
  well as the median reconstructed $m_\chi$ for the global fit based on
  minimizing $\chi^2$ of Eq.(\ref{chisq}) with $n_{\rm max} = 2$. These
  results are for our standard halo, i.e., no late infall component, without
  any cuts on the recoil energy.}
\end{figure}

In Fig.~3 
we show upper and lower bounds on the reconstructed
WIMP mass, calculated from the requirement that $\chi^2$ exceeds its minimum
by 1, for the idealized scenario where no cuts on $Q$ have been applied. We
took our default set of parameters, i.e., vanishing late infall component, and
target nuclei $X = {}^{28}$Si, $Y = {}^{76}$Ge. This figure is based on
simulating \mbox{2 $\times$ 5,000} experiments, where each experiment contains an
expected 50 events; the actual number of events is Poisson--distributed around
this expectation value.  As mentioned earlier, taking equal numbers of events
in both experiments minimizes the statistical error for fixed total number of
events. As discussed in Ref.~\cite{DMDD}, the error on the (high) moments is
not quite Gaussian; the deviation becomes larger for smaller samples. The
reason is that the high moments receive large contributions from the region of
high $Q$, where on average very few events will lie; even the region where on
average only a fraction of an event lies can contribute significantly. As a
result, most simulated experiments will underestimate these moments, while a
few (rare) experiments will overestimate them. In order to alleviate this
problem, we only include moments up to $n_{\rm max} = 2$ in our fit. Moreover,
we always show median, rather than mean, values for the (bounds on the)
reconstructed WIMP mass.

We see that the minus--first moment by itself leads to very poor bounds on
$m_\chi$. This is not surprising, since its error is dominated by the error on
the counting rate at $Q_{\rm min}$, which is determined only from the events
in the first bin. The higher moments lead to increasingly tighter bounds.
However, the higher moments are very strongly correlated. Also, the systematic
effect due to the limited event samples discussed in the previous paragraph
becomes larger for larger $n$; for example, for $n=2$ and a true $m_\chi = 1$
TeV, the median upper end of the (nominal) $1 \, \sigma$ range lies well below
1 TeV.  The lower bound on $m_\chi$ derived from the assumption of equal
scattering cross sections on protons and neutrons is similar to that derived
from the first moment, but the corresponding upper bound is significantly
worse.  Nevertheless, this estimator of $m_\chi$ helps in narrowing down the
error of the total fit, described by the upper and lower solid (red) curves.
As expected from Fig.~1 
the relative error on $m_\chi$ is
minimal for $m_\chi = \sqrt{\mX \mY}$, although the increase towards smaller
$m_\chi$ is less than expected from the behavior of the kinematical factor
$\kappa$ alone.

Unfortunately we also see that the median reconstructed $m_\chi$ starts to
deviate from the input value if $m_\chi \gsim 80$ GeV. This is a direct
consequence of the fact that the median value of the estimators of the higher
moments is too small, as discussed above. For very large $m_\chi$ the median
reconstructed WIMP mass even becomes independent of its true value; this is
true also for the upper end of the error band. This systematic shift presents
another argument in favor of imposing an upper cut $Q_{\rm max}$ on $Q$,
chosen sufficiently low that an average experiment will still have a few
events not too far below $Q_{\rm max}$.

\begin{figure}[p!] \label{fig:5050}
\begin{center}
\vspace*{-1cm}
\rotatebox{-90}{\includegraphics[width=11.cm]{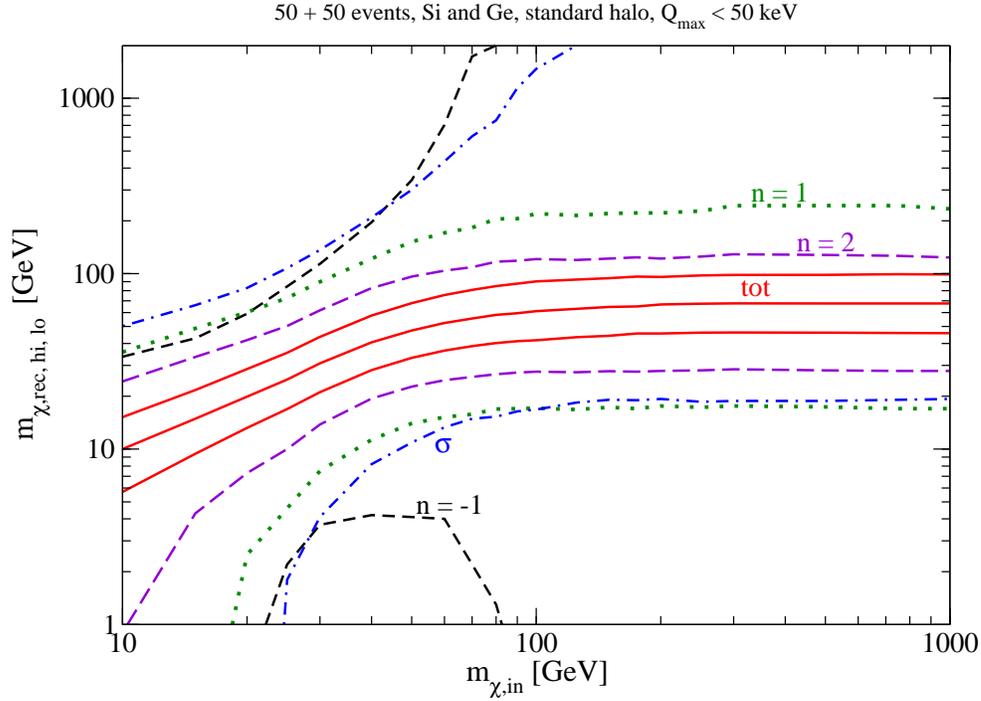}}
\end{center}
\vspace*{-1cm}
\caption{As in Fig.~3, 
  except that we have imposed the cut
  $Q_{\rm max} = 50$ keV in both experiments. Note that the average of 50
  events per experiment refers to the entire $Q$ range, i.e., the number of
  events after cuts is smaller if $Q_{\rm max,kin} > Q_{\rm max}$.}
\end{figure}
\begin{figure}[p!] \label{fig:opt}
\begin{center}
\vspace*{-.3cm}
\rotatebox{-90}{\includegraphics[width=11.cm]{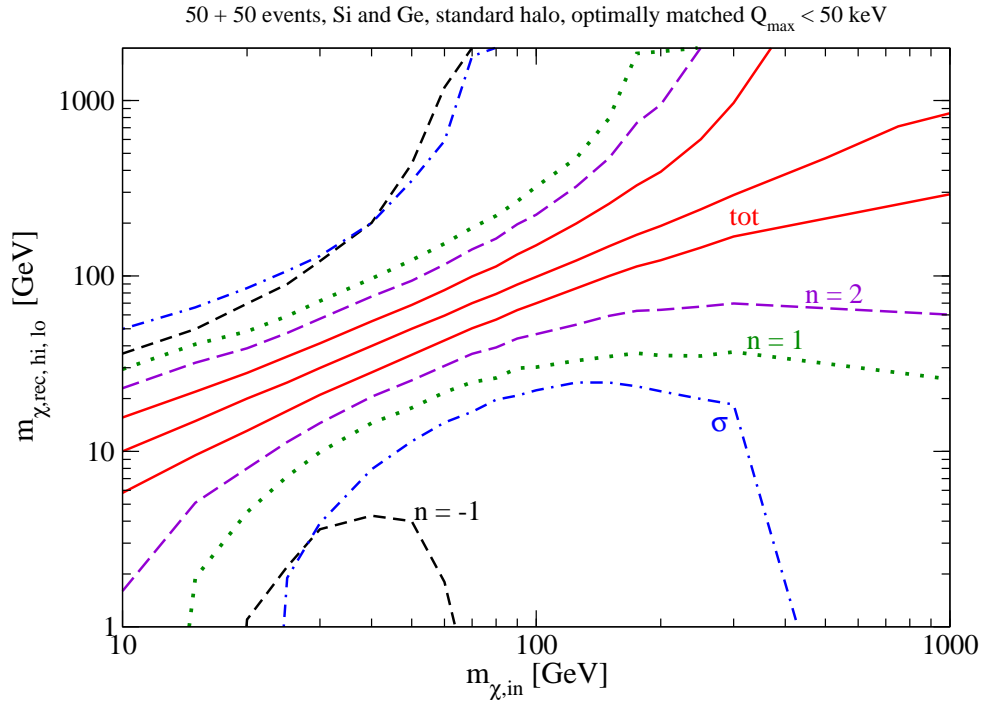}}
\end{center}
\vspace*{-1cm}
\caption{As in Fig.~4, 
  except that we have imposed the cut
  $Q_{\rm max} = 50$ keV for the Ge target. The value of $Q_{\rm max}$ for the
  Si target has been chosen such that it corresponds to the same WIMP
  velocity.}
\end{figure}

For the case at hand, with rather small event samples, this would imply
$Q_{\rm max} \simeq 50$ keV. Unfortunately Fig.~4 
shows that
simply imposing the same cut on $Q_{\rm max}$ in both targets makes the
situation significantly {\em worse}. Now the median reconstructed $m_\chi$
starts to undershoot the input value already at $m_\chi \simeq 60$ GeV, and the
asymptotic reconstructed $m_\chi$ for large input WIMP mass lies well below
100 GeV. Clearly we need to choose different $Q_{\rm max}$ values for the two
experiments if we want to get reliable results also for larger WIMP masses.

Fig.~5 
indicates that this should be possible, at least in
principle. Here we have again applied a fixed upper cut $Q_{\rm max} = 50$ keV
for the Ge experiment, but matched the cut on $Q_{\rm max}$ for the Si
experiment such that it corresponds to the same WIMP velocity:
\beq \label{match}
Q_{\rm max,\,Si} = \left( \frac {\alpha_{\rm Ge}} {\alpha_{\rm Si}} \right)^2
Q_{\rm max,\,Ge}\,,
\eeq
where $\alpha$ has been given in Eq.(\ref{eqn2105}). We see that now the
median reconstructed $m_\chi$ indeed tracks the input value even for very
large WIMP masses. It should be noted that these results still only use 50
events on average per experiment {\em before cuts}. For example, for $m_\chi =
500$ GeV Eq.(\ref{match}) with $Q_{\rm max,\,Ge} = 50$ keV implies $Q_{\rm
  max,\,Si} = 21.7$ keV, meaning that the expected number of events in the Si
experiments is only about 21. This explains why the error bands at large
values of $m_\chi$ are significantly wider here than in Fig.~3 
where no cuts on the recoil energy have been imposed. Of course, the
systematic deviation of the reconstructed $m_\chi$ from its true value makes
the error bands in Fig.~3 
largely meaningless for $m_\chi \gsim 100$
GeV.

\begin{figure}[b!] \label{fig:inout}
\begin{center}
\rotatebox{-90}{\includegraphics[width=9cm]{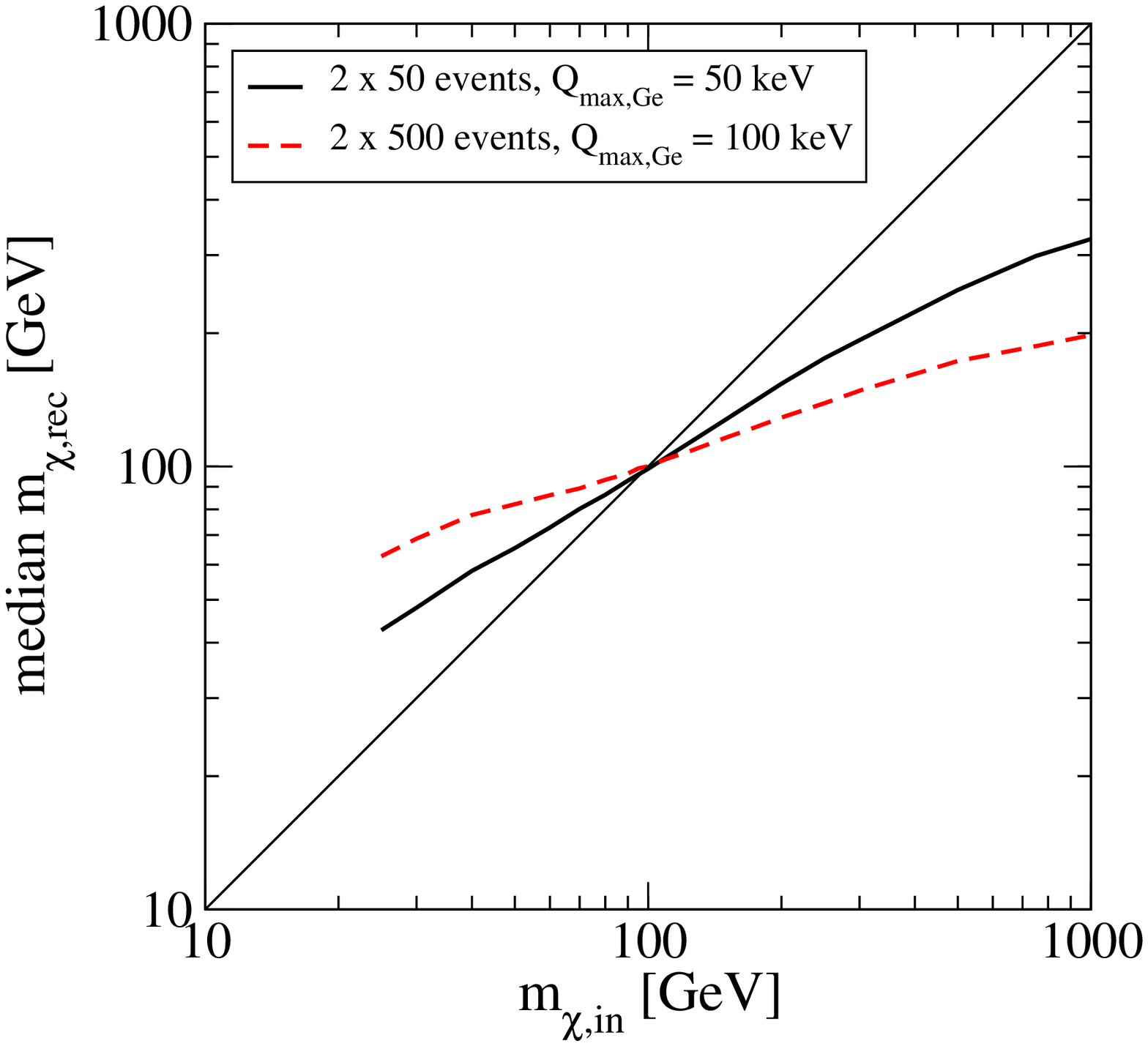}} \hspace*{-3.7cm}
\rotatebox{-90}{\includegraphics[width=9cm]{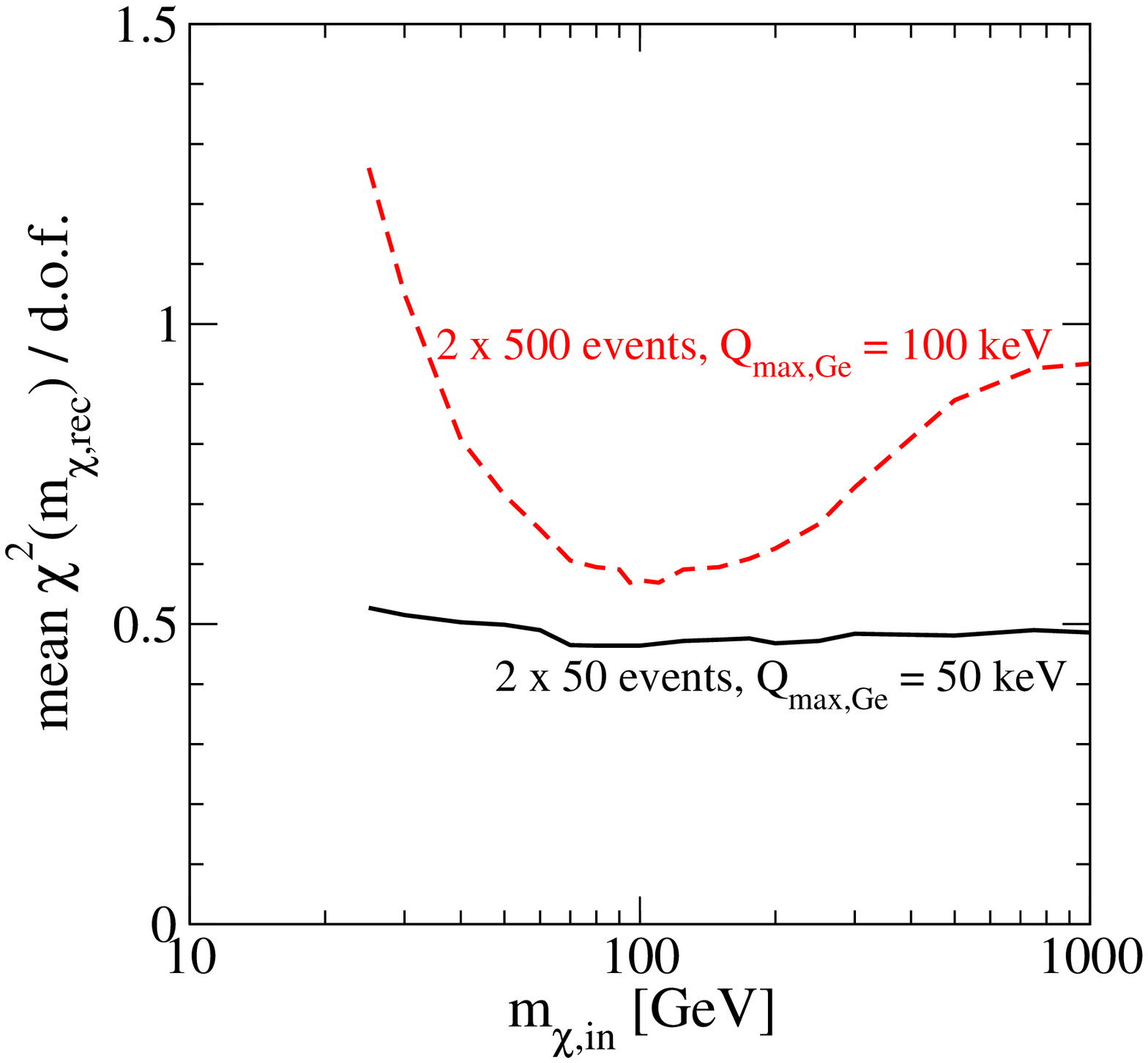} }
\end{center}
\vspace*{-1cm}
\caption{Sensitivity of the $\chi^2$ fit to the value of $m_\chi$ that is used
  as input in the $Q_{\rm max}$ matching condition (\ref{match}), assuming a
  true WIMP mass of 100 GeV, for our default choice of targets and $f_1(v)$.
  The solid (black) lines are for $2 \times 50$ events and $Q_{\rm max} = 50$
  keV, while the dashed (red) curves assume $2 \times 500$ events and $Q_{\rm
    max} = 100$ keV; here $Q_{\rm max}$ stands for the bigger of the values
  used for the two targets, the second one being fixed by Eq.(\ref{match}).
  The curves terminate at $m_{\chi,{\rm in}} = 25$ GeV since for even smaller
  WIMP masses, $Q_{\rm max,kin} < 50$ keV, making matching superfluous. The
  left frame shows the median reconstructed WIMP mass from a $\chi^2$ fit with
  $n_{\rm max} = 2$, and the right frame shows the corresponding mean value of
  $\chi^2(m_{\chi,{\rm rec}}).$}
\end{figure}

While the results in Fig.~5 
look quite impressive, they suffer
from the fact that optimal $Q_{\rm max}$ matching, as in Eq.(\ref{match}), is
only possible if $m_\chi$ is already known. The left frame of
Fig.~6 
shows that inputting the wrong value of $m_\chi$ into
Eq.(\ref{match}) will usually lead to a reconstructed WIMP mass in between
this input value and the true value. Note that the median $m_{\chi,{\rm rec}}$
as a function of $m_{\chi,{\rm in}}$ has a slope less than unity. This means
that an iteration, where one starts with some input value of $m_\chi$ and uses
the corresponding $m_{\chi,{\rm rec}}$ as new input value into
Eq.(\ref{match}), will converge ``on average''. Unfortunately our Monte Carlo
simulations show that for any one experiment, this procedure does not
necessarily converge to a well--defined $m_{\chi,{\rm rec}}$; rather, one
often ends up in an endless loop over several values of $m_{\chi,{\rm rec}}$.

An alternative is to try an algorithm for $Q_{\rm max}$ matching which is
based on the minimum value of $\chi^2$ obtained in the fit; recall that this
minimum defines the reconstructed WIMP mass. Unfortunately the right frame in
Fig.~6 
indicates that this may be difficult, at least with the
initial small statistics expected. This figure shows that the mean value of
$\chi^2(m_{\chi,{\rm rec}})$ is almost independent of the value of $m_\chi$
input into Eq.(\ref{match}) if one has only $2 \times 50$ events in the
sample. The situation looks considerably more promising with $2 \times 500$
events, and larger allowed $Q_{\rm max}$: in this case $\chi^2(m_{\chi,{\rm
    rec}})$ has a clear, if rather broad, minimum where $m_{\chi,{\rm in}}$
equals the true WIMP mass, taken to be 100 GeV in this example. 
\begin{figure}[b!] \label{fig:real}
\begin{center}
\vspace*{-1cm}
\rotatebox{-90}{\includegraphics[width=13.5cm]{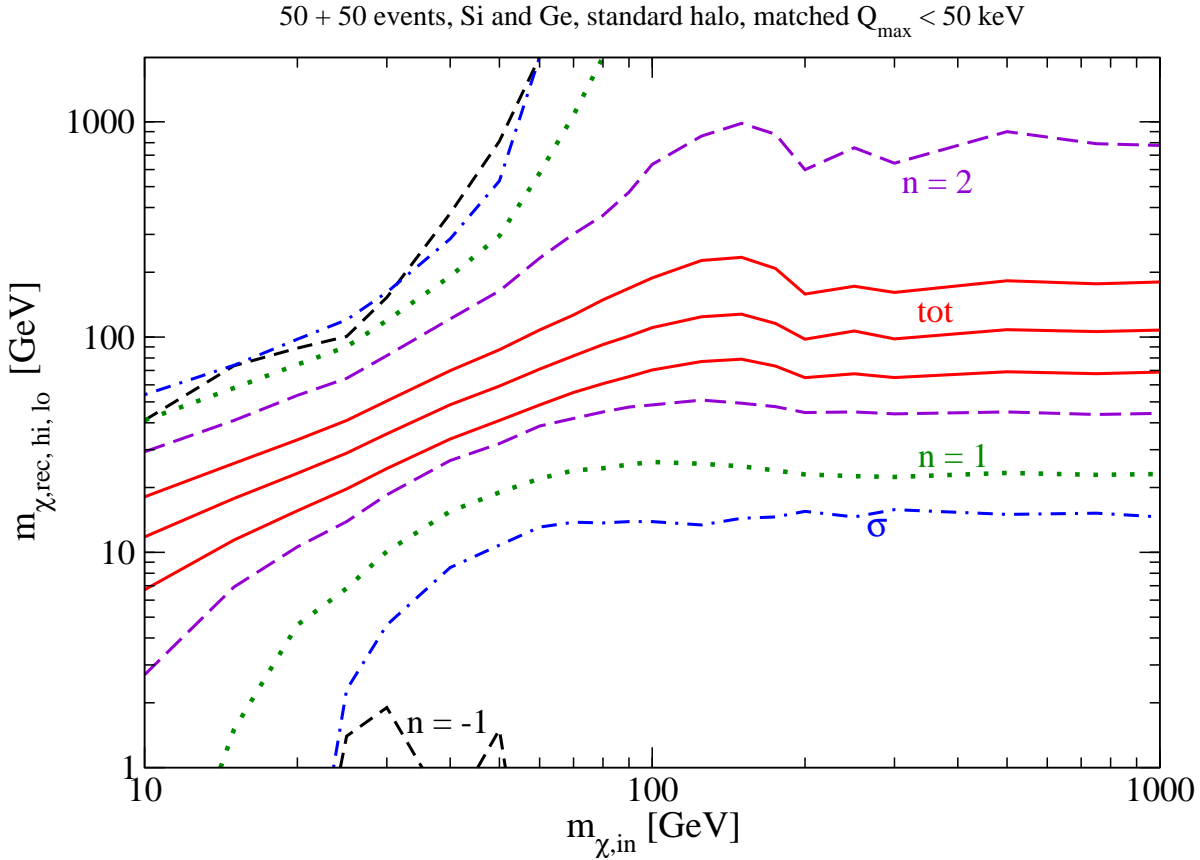}}
\end{center}
\vspace*{-1cm}
\caption{As in Fig.~4, 
  except that we have imposed the fixed cut
  $Q_{\rm max} = 50$ keV only on the Ge experiment and determined $Q_{\rm max,
  Si}$ by minimizing $\chi^2(m_{\chi,{\rm rec}})$.}
\end{figure}

Note that this minimum has mean $\chi^2(m_{\chi,{\rm rec}}) / {\rm d.o.f.}$
well below unity; the same is true for the entire solid curve. Here the number
of degrees of freedom is three, since we fit four quantities with one free
parameter. This indicates that our numerical procedure on average
over--estimates the true errors somewhat. In fact, following Ref.~\cite{DMDD},
we add the ``error on the error'' to the diagonal entries of ${\rm
  cov}(I_n,I_m)$, in order to tame non--Gaussian tails in the distribution of
measured moments of $f_1$.

One possibility is to choose the $Q_{\rm max}$ values such that
$\chi^2(m_{\chi,{\rm rec}})$ is minimal. Note that this implies a double
minimization: for fixed values of $Q_{{\rm max}, X}$ and $Q_{{\rm max},Y}$,
$m_{\chi, {\rm rec}}$ is the WIMP mass that minimizes $\chi^2(m_{\chi})$. In
an outer loop, $Q_{\rm max}$ is varied. We found that varying both $Q_{\rm
  max}$ values leads to quite misleading results, especially for larger (true)
WIMP masses and/or limited statistics. On the other hand, Fig.~4 
shows that for small $m_\chi$, $Q_{\rm max}$ matching is not really necessary;
if $m_{\chi} > \sqrt{ m_X m_Y}$, the matching condition (\ref{match}) implies
that $Q_{\rm max}$ for the lighter target should be smaller than that for the
heavier target.
\begin{figure}[b!] \label{fig:m50_100}
\begin{center}
\vspace*{-1cm}
\rotatebox{-90}{\includegraphics[width=13.5cm]{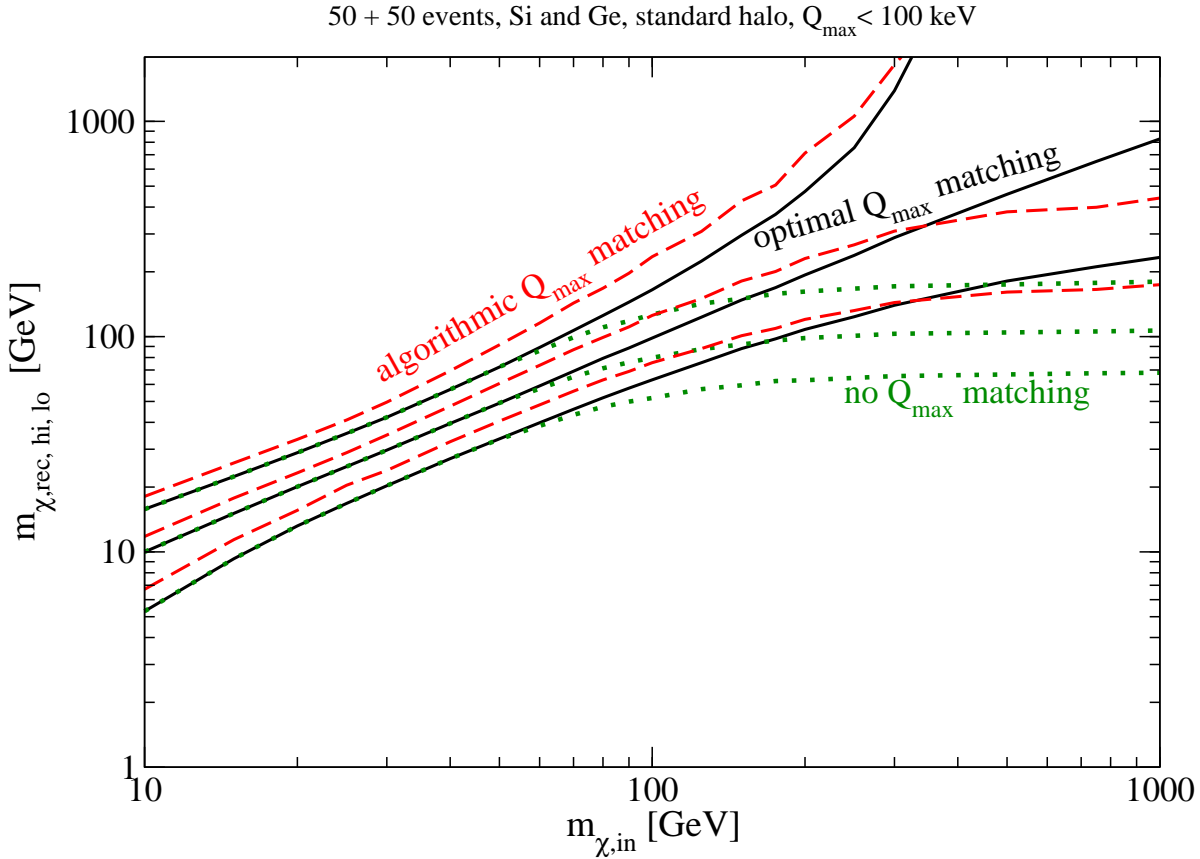}}
\end{center}
\vspace*{-1cm}
\caption{Results for the median value of the reconstructed WIMP mass as well
  as the ends of its error interval. All results are based on the combined
  fit, using Eq.(\ref{chisq}) with $n_{\rm max} = 2$. We assume Si and Ge
  targets with on average 50 events each before cuts, and took our default
  ansatz for $f_1$. The dotted (green) lines show results for $Q_{\rm max, Si}
  = Q_{\rm max, Ge} = 100$ keV, whereas the solid (black) lines have been
  obtained using Eq.(\ref{match}) with the bigger of the two $Q_{\rm max}$
  values fixed to 100 keV. Finally, the dashed (red) lines are for the case
  that $Q_{\rm max, Ge} = 100$ keV, whereas $Q_{\rm max, Si}$ has been chosen
  such that $\chi^2(m_{\chi,{\rm rec}})$ is minimal.
}
\end{figure}

In Fig.~7 
we have therefore minimized $\chi^2(m_{\chi,{\rm
    rec}})$ {\em only} for the lighter target, here Silicon. Note that we
always require $Q_{\rm max, Si} \leq Q_{\rm max, Ge}$ in this procedure; in
Fig.~7 
the latter is fixed to 50 keV. We see that this leads to a
systematic over--estimation of $m_\chi$ for small WIMP masses. For heavier
WIMPs the results are somewhat better than for the case where both $Q_{\rm
  max}$ values are simply taken to be equal, see Fig.~4. 
However,
this algorithm clearly still fails quite badly for $m_\chi \gsim 100$ GeV.
This is partly due to the low maximal value of $Q_{\rm max}$ assumed here. The
fact that the kinematic upper bound $Q_{\rm max,kin}$ of Eq.(\ref{qmaxkin})
increases with $m_\chi$ indicates that the region of large $Q$ is more
sensitive to the value of $m_\chi$ if the WIMP mass exceeds the mass of the
heaviest target nucleus.

In Fig.~8 
we have therefore increased the cut $Q_{\rm max}$ to
100 keV; of course, this cut only becomes effective once $Q_{\rm max, kin} >
100$ keV. Unlike in the previous figures, we here show results only for the
final fit; the values of $m_\chi$ derived from single observables are no
longer shown. This allows us to show results for three different choices of
$Q_{\rm max, Si}$ and $Q_{\rm max, Ge}$.
The dotted (green) curves show the median reconstructed WIMP mass and its
``$1\sigma$'' upper and lower bounds for the case where both $Q_{\rm max}$
values have been fixed to 100 keV. Due to the higher $Q_{\rm max}$ chosen
here, this works for considerably larger WIMP masses than in the corresponding
Fig.~4, 
where both $Q_{\rm max}$ values had been fixed to 50 keV.
However, for large $m_\chi$ the median reconstructed WIMP mass is still only
slightly above 100 GeV.
The solid (black) lines show results for the case that perfect $Q_{\rm max}$
matching has been applied, using Eq.(\ref{match}). Comparison with
Fig.~5 
shows that increasing $Q_{\rm max}$ actually slightly increases the width of
the errors on the reconstructed WIMP mass. The reason is that $Q_{\rm max}$ is
now so large that the median values of the estimators for the (generalized)
moments of $f_1$ fall somewhat below the true values. Nevertheless the results
for this optimal $Q_{\rm max}$ matching remain very encouraging.

Most importantly, our simple algorithm of fixing $Q_{\rm max, Ge}$, in this
case to 100 keV, and determining $Q_{\rm max, Si}$ by minimizing
$\chi^2(m_{\chi,{\rm rec}})$ now also seems to work reasonably well for WIMP
masses up to $\sim 500$ GeV. For $m_\chi \lsim 100$ GeV the median WIMP mass
determined in this way again over--estimates its true value by 15 to 20\%;
however, the median value of the ``$1\sigma$'' lower bound lies below the true
WIMP mass for all values of $m_\chi$. Similarly, the median ``$1\sigma$''
upper bound now lies well above the true WIMP mass even for $m_\chi = 1$ TeV,
in sharp contrast to the results shown in Fig~7. 

\begin{figure}[t!] \label{fig:arxe}
\begin{center}
\vspace*{-1cm}
\rotatebox{-90}{\includegraphics[width=11.5cm]{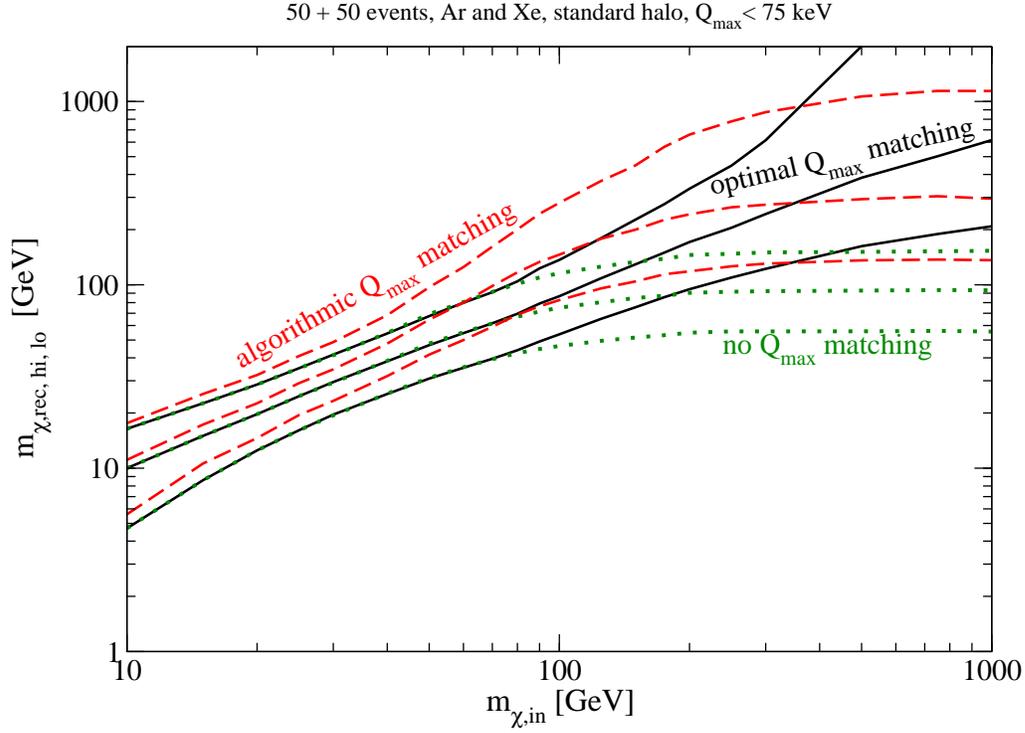}}
\end{center}
\vspace*{-1cm}
\caption{As in Fig.~8, 
  except for ${}^{40}$Ar and ${}^{136}$Xe
  targets, and with both $Q_{\rm max}$ values restricted to not exceed 75 keV.
}
\end{figure}
\begin{figure}[h!] \label{fig:li}
\begin{center}
\rotatebox{-90}{\includegraphics[width=11.5cm]{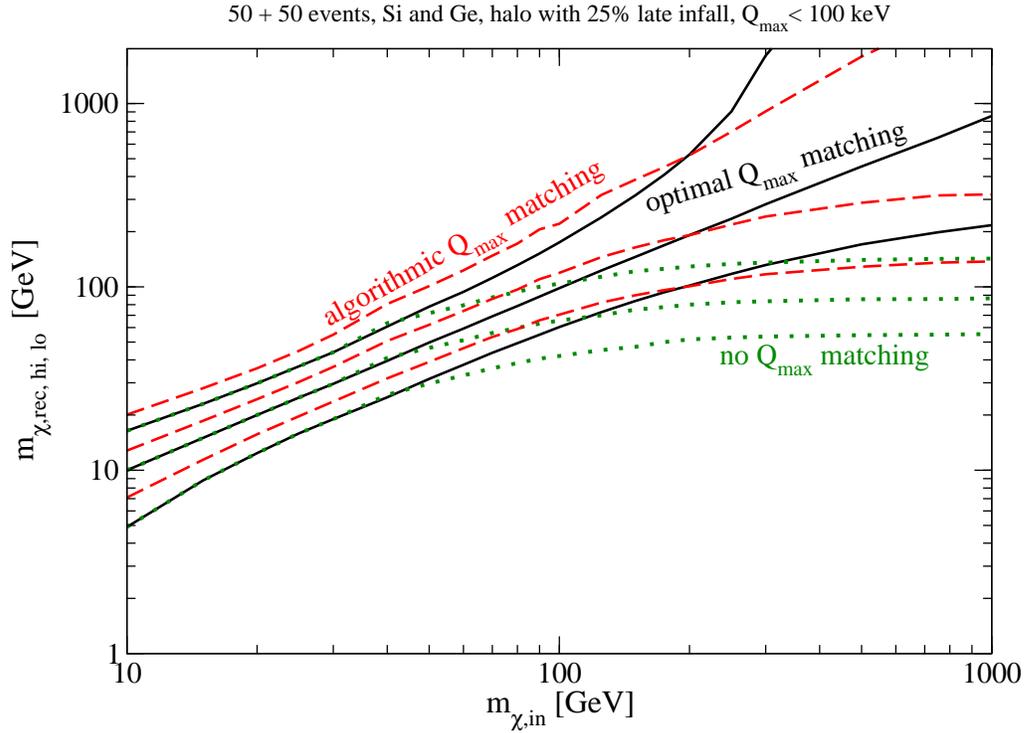}}
\end{center}
\vspace*{-1cm}
\caption{As in Fig.~8, 
  except that we allowed a 25\% late
  infall component in the local WIMP flux, by setting $N_{\rm l.i.} = 0.25$ in
  Eq.(\ref{f1}).  }
\end{figure}

We had argued earlier, based on the results of Fig.~1, 
that Si
and Ge is likely to be close to the optimal choice among the target nuclei
currently being employed. In Fig.~9 
we back this up by showing
results analogous to those of Fig.~8, 
but with Argon and Xenon
targets. Since the elastic form factor of ${}^{136}$Xe has a zero near 95 keV,
we have lowered the upper bound on $Q_{\rm max}$ to 75 keV; larger values
would start to probe the very sparsely populated region near the zero, leading
to too small median values of the estimated moments, whereas smaller values
would lead to even worse sensitivity to large WIMP masses. We see that, except
for small WIMP masses, the results are clearly worse than those shown in
Fig.~8. 

The starting point of our discussion was that we wanted to devise ways to
estimate the WIMP mass which are independent of any assumptions on the WIMP
velocity distribution $f_1$. In order to test this, we have simulated Si and
Ge experiments, allowing 25\% of the local WIMP flux to come from a ``late
infall'' component, i.e., we fixed $N_{\rm l.i.} = 0.25$ in Eq.(\ref{f1}).  The
results are shown in Fig.~10. 
We see that the results are very
similar to those with $N_{\rm l.i.}=0$ shown in Fig.~8 
if
optimal $Q_{\rm max}$ matching (\ref{match}) is used. On the other hand, the
results for non--optimal choices of $Q_{\rm max}$ get somewhat worse. This is
true both for the simple choice $Q_{\rm max,Si} = Q_{\rm max, Ge}$, where
significant deviations set in for lower values of the WIMP mass, and for our
algorithm of determining $Q_{\rm max, Si}$ by minimizing $\chi^2(m_{\chi,{\rm
    rec}})$. In the latter case, the systematic difference between median
reconstructed and true WIMP mass is larger than for $N_{\rm l.i.}=0$, both for
small and for large $m_\chi$. The reason for this degradation is that
introducing a large late--infall component, corresponding to WIMPs with
velocity about three times larger than the mean velocity of the shifted
Gaussian component, increases the number of events at large recoil energy.
Hence correct $Q_{\rm max}$ matching becomes more important. Note, however,
that the true $m_\chi$ always lies within the median limits of the
``$1\sigma$'' error interval estimated from our algorithmic $Q_{\rm max}$
matching.

\begin{figure}[t!] \label{fig:500}
\begin{center}
\vspace*{-1cm}
\rotatebox{-90}{\includegraphics[width=13.5cm]{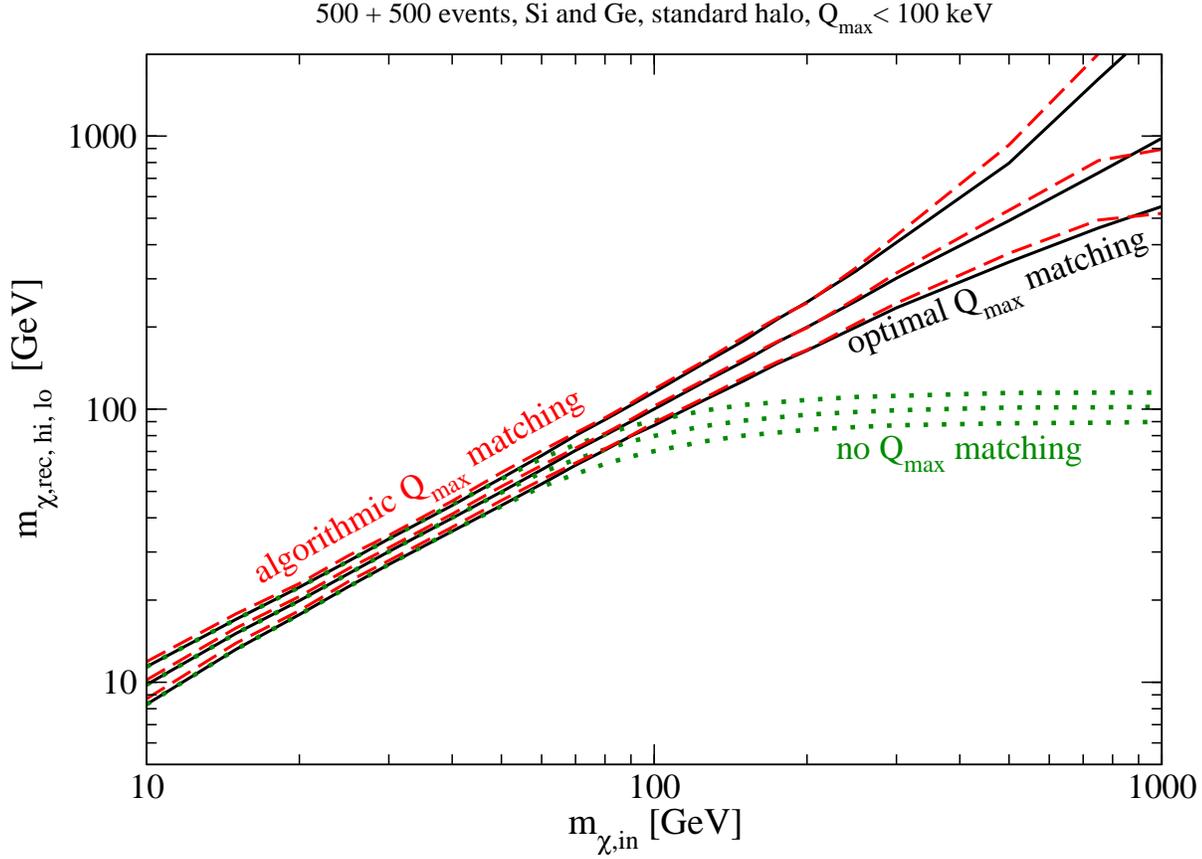}}
\end{center}
\vspace*{-1cm}
\caption{As in Fig.~8, 
except for $2 \times 500$ events before
cuts. }
\end{figure}

So far we have assumed that each experiment ``only'' has an exposure
corresponding to 50 events before cuts. In Fig.~11 
we raise this
number by a factor of 10. Not surprisingly, all error bars shrink by a factor
$\gsim~3$ compared to the situation of Fig.~8. 
The error
interval also remains approximately symmetric out to much larger values of
$m_\chi$. However, the larger number of events does not significantly change
the median reconstructed $m_\chi$ if one simply fixes both $Q_{\rm max}$
values to 100 keV. This is not surprising: for large $m_\chi$ this implies
that $v_{2,{\rm Si}}$ and $v_{2,{\rm Ge}}$ in the definition (\ref{momnew}) of
the generalized moments, and hence the moments themselves, are quite
different. Hence the estimators of these moments will not agree if the true
WIMP mass is used. On the other hand, our algorithm for fixing $Q_{\rm max,
  Ge}$ now seems to perform very well over the entire range of WIMP masses
shown. In particular, the median reconstructed WIMP mass now overshoots its
true value by only a few percent for $m_\chi \lsim 100$ GeV, and remains close
to the true value even at $m_\chi = 1$ TeV. Unfortunately we will see shortly
that for large WIMP masses our algorithm still has some problems even in this
case.

The problem lies in the distribution of the reconstructed WIMP masses in the
simulated experiments. This distribution is supposed to be characterized by
the error intervals shown in Figs.~3--5 
and 7--11. 
In order to see how well this works, we
introduce the quantity
\beq \label{delm}
\renewcommand{\arraystretch}{2.5}
\delta m = \left\{ \begin{array}{l c l}
\displaystyle
1 + \frac {m_{\chi, {\rm lo1}} - m_\chi } 
{m_{\chi, {\rm lo1}} - m_{\chi, {\rm lo2} } }\, , & ~~~~~~ &
{\rm if} \ m_\chi \leq m_{\chi,{\rm lo1}}\, ; \\
\displaystyle
\frac {m_{\chi, {\rm rec}} - m_\chi } 
{m_{\chi, {\rm rec}} - m_{\chi, {\rm lo1} } }\, , & &
{\rm if} \ m_{\chi,{\rm lo1}} < m_{\chi} < m_{\chi, {\rm rec}}\, ; \\ 
\displaystyle
\frac {m_{\chi, {\rm rec}} - m_\chi } 
{m_{\chi, {\rm hi1}} - m_{\chi, {\rm rec} } }\, , & &
{\rm if} \ m_{\chi,{\rm rec}} < m_{\chi} < m_{\chi, {\rm hi1}}\, ; \\ 
\displaystyle
\frac {m_{\chi, {\rm hi1}} - m_\chi } 
{m_{\chi, {\rm hi2}} - m_{\chi, {\rm hi1} } } - 1 \, , & &
{\rm if} \ m_\chi \geq m_{\chi,{\rm hi1}}\, .
\end{array} \right.
\eeq
\begin{figure}[b!] \label{fig:del50}
\begin{center}
\vspace*{-1cm}
\rotatebox{-90}{\includegraphics[width=13.5cm]{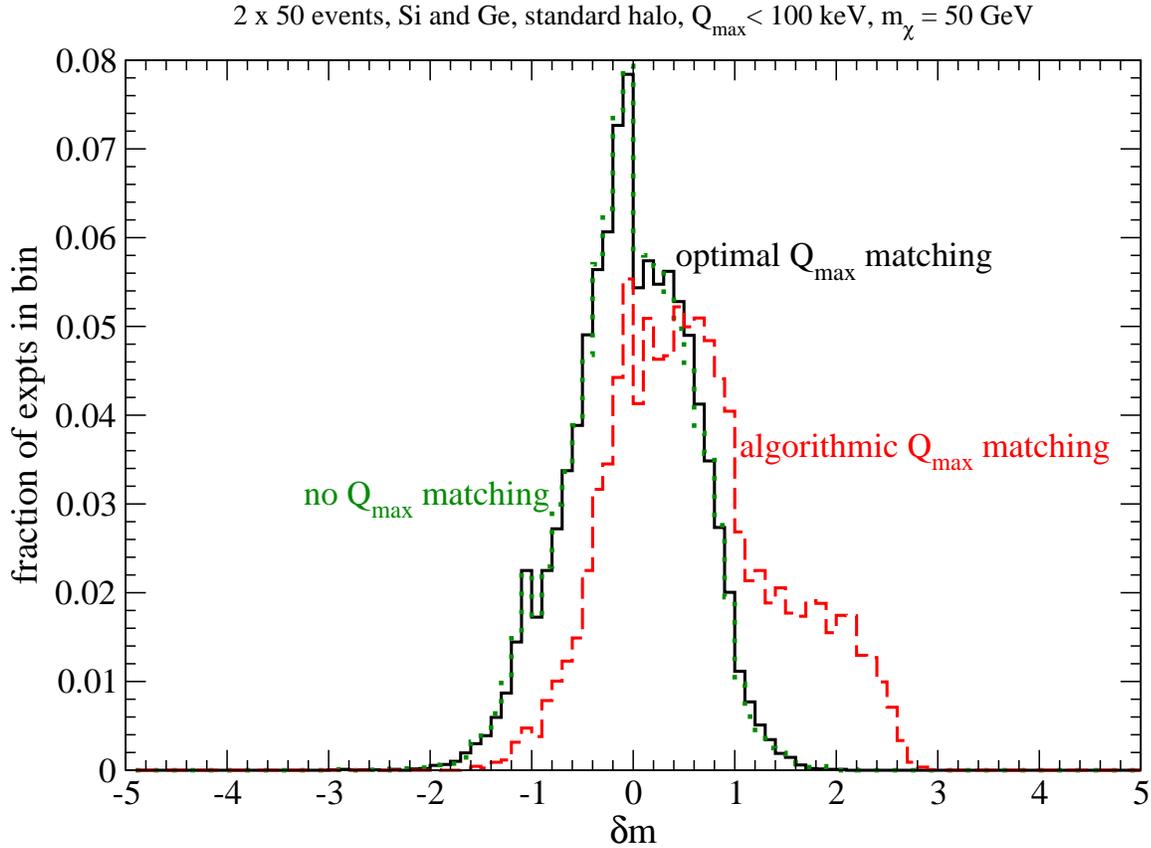}}
\end{center}
\vspace*{-1cm}
\caption{Normalized distribution of the variable $\delta m$ defined in
  Eq.(\ref{delm}) for 5,000 simulated experiments, for true WIMP mass
 $m_\chi = 50$ GeV. The other parameters, are as in Fig.~8. 
The solid (black) histogram shows results for perfect $Q_{\rm max}$ matching
 with $Q_{\rm max} \leq 100$ keV, based on Eq.(\ref{match}), whereas the
 dotted (green) histogram is for fixed equal $Q_{\rm max}$ values of 100
 keV. The dashed (red) histogram shows results when $Q_{\rm max, Si}$ is
 determined by minimizing $\chi^2(m_{\chi,{\rm rec}})$.
 }
\end{figure}
Here $m_\chi$ is the true WIMP mass, $m_{\chi,{\rm rec}}$ its reconstructed
value, $m_{\chi, {\rm lo1(2)}}$ is the ``$1\ (2)\, \sigma$'' lower bound
satisfying $\chi^2(m_{\chi,{\rm lo(1,2)}}) = \chi^2(m_{\chi,{\rm rec}}) + 1\ 
(4)$, and $m_{\chi, {\rm hi1(2)}}$ are the corresponding ``$1\ (2)\,\sigma$''
upper bounds.  In the limit of purely Gaussian errors, where $\chi^2$ of
Eq.(\ref{chisq}) is simply a parabola, $(\delta m)^2$ would itself be a
$\chi^2$ variable, measuring the difference between the true and the
reconstructed WIMP mass in units of the error of the reconstruction. However,
we saw earlier that the error intervals are often quite asymmetric. Similarly,
the distance between the ``$2\sigma$'' and ``$1\sigma$'' limits can be quite
different from the distance between the ``$1\sigma$'' limit and the central
value. The definition (\ref{delm}) takes these differences into account, and
also keeps track of the sign of the deviation: if the reconstructed WIMP mass
is larger (smaller) than the true one, $\delta m$ is positive (negative).
Moreover, $|\delta m| \leq 1 \ (2)$ if and only if the true WIMP mass lies between the
``experimental'' $1 \ (2)\ \sigma$ limits.

In Fig.~12 
we show the distribution of $\delta m$ calculated from
5,000 simulated experiments, assuming a rather small WIMP mass, $m_\chi = 50$
GeV. The other parameters have been fixed as in Fig.~8. 
In
this case simply fixing both $Q_{\rm max}$ values to 100 keV still works fine,
since the kinematic maximum values of $Q$ lie only slightly above 100 keV (at
122 keV for Si and 132 keV for Ge). The distributions for fixed $Q_{\rm max}$,
or for optimal $Q_{\rm max}$ matching, look somewhat lopsided, since the error
interval is already asymmetric, with $m_{\chi,{\rm hi1}} - m_{\chi,{\rm rec}}
> m_{\chi,{\rm rec}} - m_{\chi,{\rm lo1}}$. As a result, negative values of
$\delta m$ have a larger denominator than positive values, hence the
distribution is narrower for $\delta m < 0$. These distributions also indicate
that our errors are indeed over--estimated, since nearly 90\% of the simulated
experiments have $|\delta m| \leq 1$; we remind the reader that a usual $1
\sigma$ interval should only contain some 68\% of the experiments. 

We saw in Fig.~8 
that our algorithm for determining $Q_{\rm max,
  Si}$ tends to overestimate the WIMP mass if the latter is small. This is
reflected by the dashed (red) histogram in Fig.~12, 
which has
significantly more entries at positive values than at negative values.
Moreover, this histogram is rather flat between $\delta m = 1 $ and $2$. Since
our algorithm is based on a double minimization of $\chi^2$ defined in
Eq.(\ref{chisq}), it is not very surprising that the resulting final
$\chi^2(m_{\chi,{\rm rec}})$ values are distributed quite differently from
what one finds after a single minimization step. Nevertheless it is
  reassuring that some 70\% of the simulated experiments lead to $|\delta m|
  \leq 1$.

\begin{figure}[b!] \label{fig:del200}
\begin{center}
\rotatebox{-90}{\includegraphics[width=9cm]{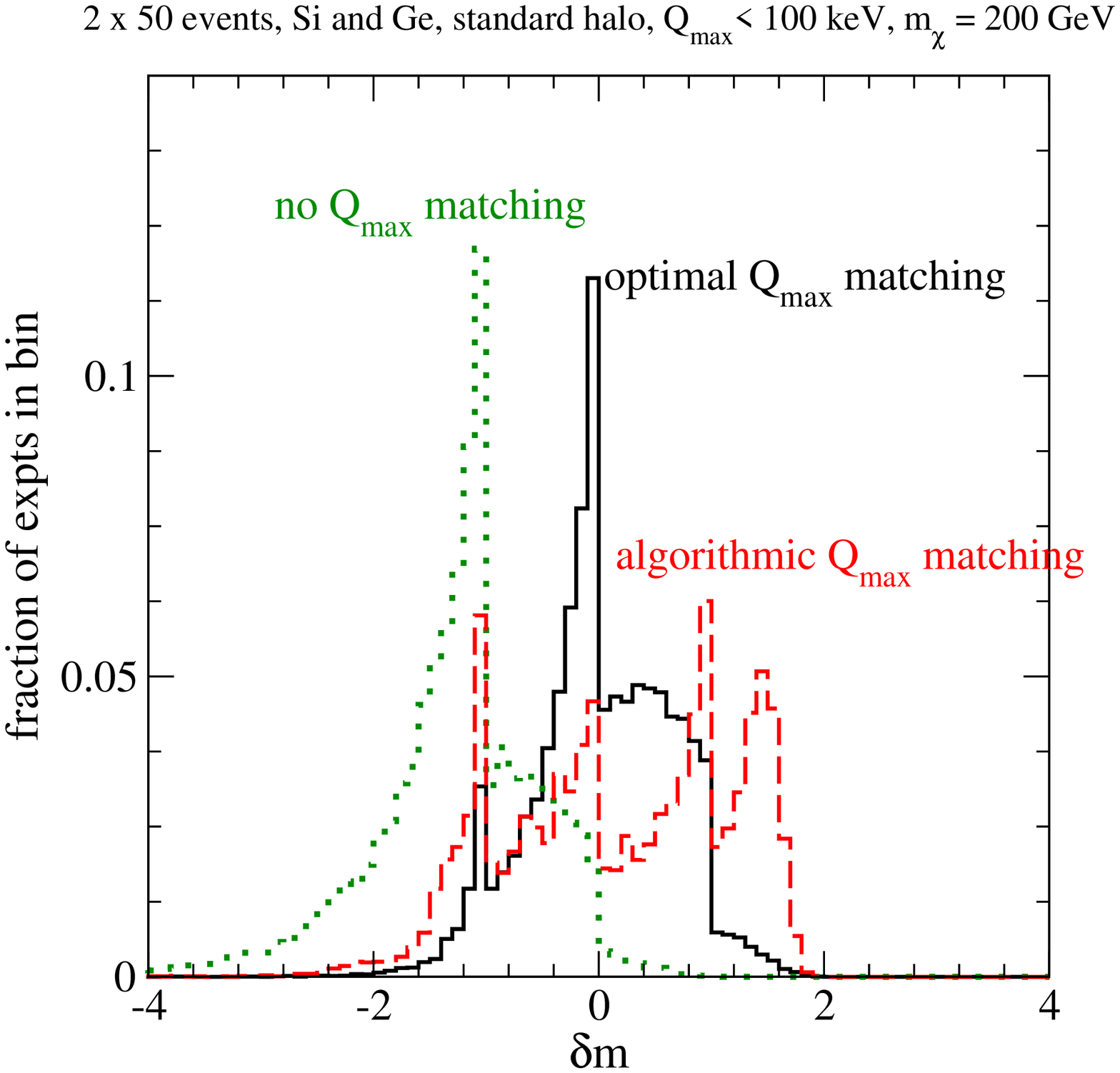}}
\rotatebox{-90}{\includegraphics[width=9cm]{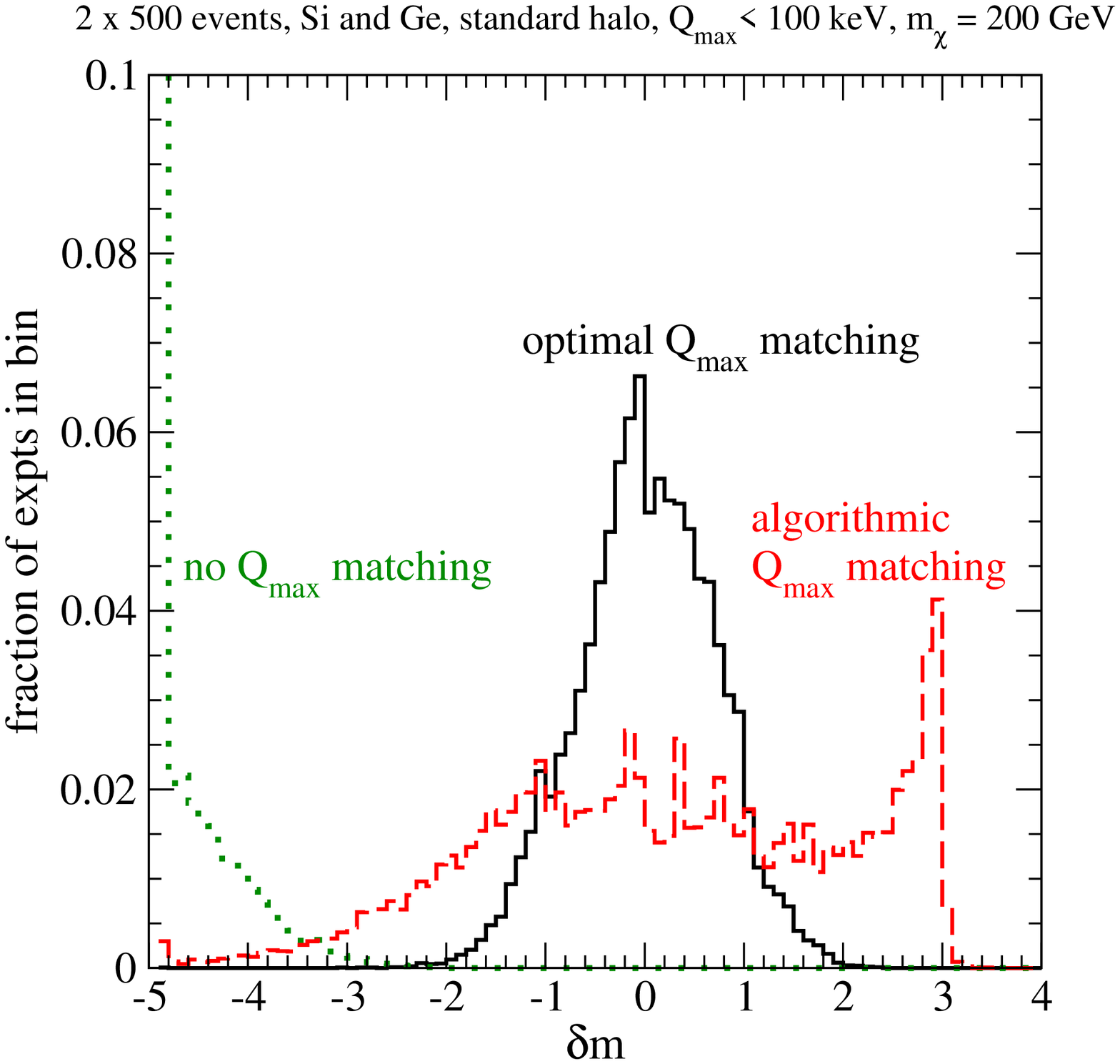} }
\end{center}
\vspace*{-1cm}
\caption{Distribution of $\delta m$ defined in Eq.(\ref{delm}) calculated from
5,000 simulated experiments. Parameters and notations are as in
Fig.~12, 
except that we have increased the WIMP mass to 200
GeV. In the right frame we have in addition increased the average number of
events (before cuts) in each experiment from 50 to 500. Note that the bins at
$\delta m = \pm 5$ are overflow bins, i.e., they also contain all experiments
with $|\delta m| > 5$.}
\end{figure}

Unfortunately Fig.~13 
shows that the situation becomes much less
favorable at the larger WIMP mass of 200 GeV. Here we show results with 50
(left) and 500 (right) events per experiment (before cuts), with the other
parameters as in Fig.~12. 
We see that for optimal $Q_{\rm max}$
matching a large majority of the simulated experiments still satisfy $|\delta
m| \leq 1$. The fact that this fraction decreases from nearly 90\% for the
smaller events samples to about 85\% for the larger samples indicates that our
error estimates become a bit more reliable for larger event
numbers.\footnote{Assuming, unrealistically, that there are 50,000 events in
  each experiment, we find that only about 75\% of all experiments have
  $|\delta m| \leq 1$, indicating a very slow approach to the Gaussian value
  of about 68\%.} Note also that the secondary peak at $\delta m \simeq -1$
is due to the change of denominator in the definition (\ref{delm}).

We already saw in Figs.~8 
and 11 
that setting
$Q_{\rm max, Si} = Q_{\rm max, Ge} = 100$ keV significantly under--estimates
the true WIMP mass if the latter exceeds 100 GeV. This is borne out by
Figs.~13. 
Since the statistical error decreases with increasing
number of events, $\delta m$ is much smaller in the right frame; in fact, most
of the simulated experiments now fall in the ``underflow bin'' $\delta m =
-5$, which also contains all experiments giving even smaller values. In other
words, most of the simulated experiments would give a reconstructed WIMP mass
more than five estimated standard deviations below the true value, if no
$Q_{\rm max}$ matching is used.

Unfortunately the error estimates resulting from our algorithm of determining
$Q_{\rm max, Si}$ by minimizing $\chi^2(m_{\chi,{\rm rec}})$ are also not very
reliable in this case. For the smaller event sample, we find that about 58\%
of the simulated experiments yield $|\delta m| \leq 1$, while nearly all
experiments give $|\delta m| \leq 2$. While these numbers are not so different
from the corresponding Gaussian predictions, the distribution of $\delta m$ is
clearly highly non--Gaussian in this case. Just as in the scenario without
$Q_{\rm max}$ matching the error estimates actually become less reliable with
increasing event samples. If both experiments contain on average 500 events
each, less than 40\% of the experiments have $|\delta m| \leq 1$, while more
than 30\% of the simulations yield $|\delta m| \geq 2$. In fact, the most
likely value of $\delta m$ is now close to 3. In view of these observations,
the fact that the median $\delta m$ is close to zero, so that the median
reconstructed WIMP mass is close to the true value as already shown in
Fig.~11, 
seems almost accidental. Recall also that the nominal ``$1
\sigma$'' uncertainty of the reconstructed WIMP mass still amounts to about 40
GeV in this case. This means that the most likely value of $m_{\chi,{\rm
    rec}}$ predicted by our algorithm exceeds the true value by more than 100
GeV. This clearly leaves some room for improvements. The fact that optimal
$Q_{\rm max}$ matching continues to give good results for both the
reconstructed WIMP mass and its error indicates that better data--based
algorithms might very well exist.

\section{Summary and Conclusions}

In this paper we described methods to determine the mass of a Weakly
Interacting Massive Particle detected ``directly'', i.e., through the recoil
energy deposited in a detector by the recoiling nucleus after a WIMP scattered
elastically off this nucleus. Our methods are model--independent in the sense
that they do not need any assumption about the WIMP velocity distribution. The
price one has to pay for this is that one will need positive signals in at
least two different detectors, employing different target nuclei. 

Our methods are based on our earlier work \cite{DMDD} on reconstructing the
WIMP velocity distribution, which we briefly reviewed in Sec.~2. In this
earlier work, which was based on results from a single (simulated) experiment,
the WIMP mass $m_\chi$ was an input. In Sec.~3 we showed how one can determine
$m_\chi$ by equating results obtained by different experiments. Here the
moments of the velocity distribution function are particularly useful, since
all events in the sample contribute to any given moment, leading to relatively
low statistical uncertainties. We also described a method for determining
$m_\chi$ that can be used if the ratio of WIMP scattering cross sections on
protons and neutrons is known; this is true, for example, for the
spin--independent scattering of supersymmetric neutralinos, where these two
cross sections are nearly equal. We also showed how to combine these methods
using a $\chi^2$ fitting procedure.

Sec.~4 was devoted to a detailed numerical analysis of our methods. We saw
that, assuming the sizes of the event samples are fixed, the statistical
errors will be smaller for larger mass difference between the two target
nuclei. In practice experiments with heavier targets will accumulate more
events, assuming equal exposure, at least if the spin--independent
contribution to the scattering cross section dominates. However, the number of
useful events (after cuts, preferably in an almost background--free energy
range) also depends on other factors, besides the masses of the target nuclei.

In our discussion we saw that, for WIMP masses exceeding 100 GeV or so, the
maximal recoil energy $Q_{\rm max}$ of accepted signal events plays a crucial
role. Existing experiments have $Q_{\rm max} \leq 100$ keV. If both targets
used fixed $Q_{\rm max}$ values of this order or even smaller, a significant
systematic error on the extracted WIMP mass results. In principle this problem
can be solved by matching the $Q_{\rm max}$ values of the two experiments. The
problem is that perfect matching requires prior knowledge of the WIMP mass. We
tried two algorithms to overcome this problem. Determining $Q_{\rm max}$ of
one experiment iteratively should converge ``on average'', but in a given
experiment often leads to an endless loop, rather than a specific value of
$Q_{\rm max}$; this problem is particularly severe for small event samples. On
the other hand, determining $Q_{\rm max}$ of the experiment with the lighter
target nucleus by minimizing $\chi^2$ also with respect to this quantity
over--estimates the WIMP mass if it is small, and leads to unreliable error
estimates if the WIMP mass is larger, the problem becoming worse with
increasing event samples. However, the fact that optimal $Q_{\rm max}$
matching works well in all cases, for both the median reconstructed WIMP mass
and its error (which tends to be over--estimated by our expressions), gives us
hope that a better algorithm for $Q_{\rm max}$ matching can be found which
only relies on the data. One possibility that might be worth exploring is to
employ a combination of an iterative procedure and a second $\chi^2$
minimization, where the latter is used only if the former does not converge to
a well--defined value of the reconstructed WIMP mass. 

We also found that imposing a cut $Q_{\rm max}$ may actually be beneficial for
small event samples. This is related to our earlier observation \cite{DMDD}
that a typical experiment will under--estimate the higher moments of $f_1$,
which receive significant contributions from recoil energies where only a
fraction of an event is expected to occur in a given experiment. This problem
becomes more acute for heavier target nuclei, since they have softer form
factors. In particular, using Xenon rather than Germanium does not improve the
determination of $m_\chi$, since the spin--independent elastic form factor of
Xenon as predicted by the Woods--Saxon ansatz vanishes for $Q\simeq 95$ keV.
In contrast, the lower (threshold) energy of the experiment does not seem to
be very important, if it can be pushed down to values near 3 keV or less.

Our analysis is idealized in that we ignore backgrounds, systematic
uncertainties as well as the finite energy resolution. The relative error on
the recoil energy in existing experiments is small compared to our most
optimistic relative WIMP mass error estimates even with 500 events per
experiment, so ignoring it should be a good approximation. Modern methods of
discriminating between nuclear recoils and other events, combined with muon
veto and good shielding, hold out the possibility of keeping (some)
experiments nearly background--free also in future. 

In our numerical analysis we have ignored the expected annual modulation of
the WIMP flux. In practice this can be done if one simply sums all events over
(at least) one full calendar year. In principle one can also use our methods
for subsets of data collected during specific times of the year. However, at
least if the standard ``shifted Gaussian'' velocity distribution is
approximately correct, we do not expect the small annual modulation to play a
significant role, even if one compares experiments taken during different
parts of the year.

We saw that our methods work best if the WIMP mass lies in between the masses
of the two target nuclei. Even in that case the error will likely be
significantly larger than the error on $m_\chi$ from collider experiments, if
the WIMP is part of a well--motivated extension of the Standard Model of
particle physics, e.g., if it is the lightest neutralino \cite{susydm} or the
lightest $T-$odd particle in ``Little Higgs'' models \cite{little}. It will
nevertheless be crucial to determine the WIMP mass from direct and/or indirect
Dark Matter experiments as precisely as possible, in order to make sure that
the particle produced at colliders is indeed the WIMP detected by these
experiments. Once one is confident of this identification, one can use further
collider measurements to constrain the WIMP couplings. This in turn will allow
to calculate the WIMP--nucleus scattering cross section. Together with the
determination of the WIMP velocity distribution \cite{DMDD}, this will then
yield a determination of the local WIMP number density via the total counting
rate in direct detection experiments. Knowledge of the WIMP couplings will
also permit prediction of the WIMP annihilation cross section. Together with
the Dark Matter density inferred from cosmological observations, this will
allow to test our understanding of the early universe \cite{early}. A
determination of the WIMP mass from Dark Matter detection experiments is thus
a crucial ingredient in many analyses that shed light on the dark sector of
the universe.

\subsubsection*{Acknowledgments}

This work was partially supported by the Marie Curie Training Research Network
``UniverseNet'' under contract no. MRTN-CT-2006-035863, as well as by the
European Network of Theoretical Astroparticle Physics ENTApP ILIAS/N6 under
contract no. RII3-CT-2004-506222.

\appendix
\setcounter{equation}{0}
\renewcommand{\theequation}{A\arabic{equation}}
\section{Derivatives needed in the error analysis}

At the end of Sec.~2 we gave the covariance matrix of the quantities appearing
in the definition of the ${\cal R}_n$ as well as ${\cal R}_\sigma$. In
Eqs.(\ref{sigmom}) and (\ref{sigsig1}) we also gave expressions relating the
errors on the reconstructed WIMP masses to the errors on ${\cal R}_n$ and
${\cal R}_\sigma$. The only missing ingredients in the calculation of the
errors on our various estimators of $m_\chi$ are the first derivatives of
${\cal R}_n$ and ${\cal R}_\sigma$.

We begin with the former. From Eq.(\ref{eqn3201}), it can be found directly
that 
\cheqnXa{A}
\beqn
\Pp{\calRn}{r_X(Q_{{\rm min},X})} \=     \frac{2}{n}
\bfrac{  Q_{{\rm min},X}^{(n+1)/2} \IzX-(n+1) Q_{{\rm min},X}^{1/2} \InX}
      {2 Q_{{\rm min},X}^{(n+1)/2} r_X(Q_{{\rm min},X})+(n+1)
 \InX F^2_X(Q_{{\rm min},X}) }
        \non\\
 \conti ~~~~~~~~~~~~~~~~ \times 
  \bfrac{F^2_X(Q_{{\rm min},X})} {2 Q_{{\rm min},X}^{1/2} r_X(Q_{{\rm min},X})
+ \IzX F^2_X(Q_{{\rm min},X})} \calRn\, ,
\eeqn
\cheqnXb{A}
\beqn
\Pp{\calRn}{\InX} \= \frac{n+1}{n}
\bfrac{F^2_X(Q_{{\rm min},X})} {2 Q_{{\rm min},X}^{(n+1)/2}
r_X(Q_{{\rm min},X}) + (n+1) \InX F^2_X(Q_{{\rm min},X})} \calRn\, ,
    \non\\
\eeqn
and
\cheqnXc{A}
\beq
\Pp{\calRn} {\IzX} =-\frac{1}{n} \bfrac{F^2_X(Q_{{\rm min},X})}
{2 Q_{{\rm min},X}^{1/2} r_X(Q_{{\rm min},X}) 
+ \IzX F^2_X(Q_{{\rm min},X})} \calRn\, .
\eeq
\cheqnX{A}
By first exchanging $Q_{{\rm min},X}^{(n+1)/2}$ and $(n+1) \InX$ with $Q_{{\rm
    min},X}^{1/2}$ and $\IzX$, respectively, and then replacing $X$ by $Y$,
one finds
\cheqnXa{A}
\beqn
\Pp{\calRn} {r_Y(Q_{{\rm min},Y})} \=    -\frac{2}{n}
 \bfrac{ Q_{{\rm min},Y}^{(n+1)/2} \IzY - (n+1) Q_{{\rm min},Y}^{1/2} \InY}
{2 Q_{{\rm min},Y}^{(n+1)/2} r_Y(Q_{{\rm min},Y}) 
+ (n+1) \InY F^2_Y(Q_{{\rm min},Y})}
        \non\\
\conti ~~~~~~~~~~~~ \times 
\bfrac{F^2_Y(Q_{{\rm min},Y})} {2 Q_{{\rm min},Y}^{1/2} r_Y(Q_{{\rm min},Y})
+ \IzY F^2_Y(Q_{{\rm min},Y})} \calRn \, ,
\eeqn
\cheqnXb{A}
\beqn
\Pp{\calRn}{\InY} \=-\frac{n+1}{n}
\bfrac{F^2_Y(Q_{{\rm min},Y})} {2 Q_{{\rm min},Y}^{(n+1)/2} 
r_Y(Q_{{\rm min},Y})+(n+1) \InY F^2_Y(Q_{{\rm min},Y})} \calRn\, ,
    \non\\
\eeqn
and
\cheqnXc{A}
\beq
\Pp{\calRn} {\IzY} = \frac{1}{n} \bfrac{F^2_Y(Q_{{\rm min},Y})}
{2 Q_{{\rm min},y}^{1/2} r_Y(Q_{{\rm min},Y}) + \IzY F^2_Y(Q_{{\rm min},Y})} 
\calRn\, .
\eeq
\cheqnX{A}
Note that a factor ${\cal R}_n$ appears in all these expressions; this allows
to cast the final result for the error on $m_\chi$ estimated using moments of
$f_1$ into a form analogous to that in Eq.(\ref{eqn3105}) even in the presence
of non--trivial cuts on the recoil energy, with the same prefactor. Moreover,
all the $\IzX, \, \IzY, \, \InX, \, \InY$ should be understood to be computed
according to Eq.(\ref{In}) or its discretization (\ref{eqn2210}) with
integration limits $Q_{\rm min}$ and $Q_{\rm max}$ specific for that target.

Similarly, the derivatives of ${\cal R}_\sigma$ can be computed from
Eq.(\ref{Rsigma1}): 
\cheqnXa{A}
\beq \label{Rsigmadifa}
\Pp{{\cal R}_\sigma}{r_X(Q_{{\rm min},X})} = \bfrac {2 Q_{{\rm min},X}^{1/2}} 
{2 Q_{{\rm min},X}^{1/2} r_X(Q_{{\rm min},X}) + \IzX F_X^2(Q_{{\rm min},X})}
{\cal R}_\sigma\, ,
\eeq
 and
\cheqnXb{A}
\beq \label{Rsigmadifb}
\Pp{{\cal R}_\sigma}{\IzX} = \bfrac {F_X^2(Q_{{\rm min},X})}
{2 Q_{{\rm min},X}^{1/2} r_X(Q_{{\rm min},X}) + \IzX F_X^2(Q_{{\rm min},X})}
{\cal R}_\sigma\, ,
\eeq
\cheqnX{A}
$\!\!\!$
The derivatives with respect to the $Y$ variables can be obtained from
Eqs.(\ref{Rsigmadifa}) and ({\ref{Rsigmadifb}}) by simply changing $X \lto Y$ everywhere
and changing the overall plus signs to minus signs.

\end{document}